\documentclass[a4paper,11pt]{article}
\pdfoutput=1 

\usepackage{jcappub} 

\usepackage[T1]{fontenc} 
\usepackage{graphicx}
\usepackage[dvipsnames]{xcolor}
\usepackage{hyperref}
\usepackage{multirow}
\usepackage{cleveref}
\usepackage{float}
\usepackage{subfig}

\usepackage{natbib}
\setcitestyle{square, numbers}

\title{\boldmath Neural Networks for cosmological model selection and feature importance using Cosmic Microwave Background data}

\author[a,1]{I. Ocampo,}
\author[b]{G. Cañas-Herrera,}
\author[a]{S. Nesseris}

\affiliation[a]{Instituto de F\'isica Te\'orica UAM-CSIC, Universidad Auton\'oma de Madrid, Cantoblanco, 28049 Madrid, Spain}
\affiliation[b]{ESTEC - European Space Agency, Keplerlaan 1, 2201 AZ Noordwijk, The Netherlands}

\emailAdd{indira.ocampo@csic.es}
\emailAdd{Guadalupe.CanasHerrera@esa.int}
\emailAdd{savvas.nesseris@csic.es}

\abstract { 
The measurements of the temperature and polarisation anisotropies of the Cosmic Microwave Background (CMB) by the ESA Planck mission have strongly supported the current concordance model of cosmology. However, the latest cosmological data release from ESA Planck mission still has a powerful potential to test new data science algorithms and inference techniques. In this paper, we use advanced Machine Learning (ML) algorithms, such as Neural Networks (NNs), to discern among different underlying cosmological models at the angular power spectra level, using both temperature and polarisation Planck 18 data. We test two different models beyond $\Lambda$CDM: a modified gravity model: the Hu-Sawicki model, and an alternative inflationary model: a feature-template in the primordial power spectrum. Furthermore, we also implemented an interpretability method based on \texttt{SHAP} values to evaluate the learning process and identify the most relevant elements that drive our architecture to certain outcomes. We find that our NN is able to distinguish between different angular power spectra successfully for both alternative models and $\Lambda$CDM. We conclude by explaining how archival scientific data has still a strong potential to test novel data science algorithms that are interesting for the next generation of cosmological experiments.}

\begin{document}
\maketitle
\flushbottom

\section{Introduction}
\label{sec:intro}

The Cosmic Microwave Background (CMB) is a remnant radiation from after the Hot Big Bang, and its detailed observations are crucial to the era of precision cosmology \cite{barreiro2010overview}. The CMB consists of photons that were emitted around 380 000 years after the Big Bang, at a time when the Universe had cooled enough for neutral atoms to form (at a redshift about $z\sim1100$). Prior to this period, the Universe was a hot and dense plasma, where photons were continuously scattering with electrons. For this reason, the CMB represented the photon-electron last scattering surface of the moment when both types of particles were coupled to each other. It contains important information about the underlying cosmological model that best explains the latest observations of our Universe \cite{baumann2009tasi, hu2008lecture}.

The Planck satellite, launched by the European Space Agency in 2009 \cite{collaboration2014planck, ade2016planck, aghanim2020planck}, is a reference in modern cosmology thanks to its exquisite measurements of the temperature and polarisation anisotropies of the CMB \cite{akrami2020planck}. The CMB radiation is highly isotropic with a temperature of approximately 2.7 K, as first studied by the COBE satellite \cite{smoot1992structure}, and it is predicted to present small anisotropies showing under and overdensed regions in the primordial universe. In fact, the NASA WMAP satellite \cite{spergel2003first, hinshaw2013nine}, which followed on COBE, and the ESA Planck satellite \cite{ade2011planck}, measured the temperature and polarization anisotropies to an exceptionally precise level. The discovery of these anisotropies presented in the CMB introduced a revolutionary approach to understand how our Universe works, and how the initial seeds that populated the very early Universe evolved to form the current Large-Scale Structure (LSS) we observed in the Universe. In general, we study the underlying cosmological model by comparing observable predictions, such as the angular power spectrum decomposition of the temperature and polarisation anisotropies, against real CMB data. These observations enabled to impose stringent constraints on the underlying physical model, called the standard cosmological concordance model, or simply, the $\Lambda$CDM (with the best fit values in table \autoref{tab:cosmo_param}). This model assumes that the Universe's accelerated expansion can be explained in terms of a Cosmological constant in the dynamic equations ($\Lambda$, popularly associated to Dark Energy), that the main energy density component is Cold Dark Matter (CDM), and that the primordial density perturbations in the early Universe were almost Gaussianly-distributed, compatible with the predictions of the vanilla single-field inflation model. On top, CMB data has given powerful insights on alternative inflationary models \cite{akrami2020planck}, the nature of dark matter and dark energy \cite{ade2016planck, galloni2023unraveling, scott2016information}, and the homogeneity of our Universe \cite{Planck:2019evm}. 

\begin{table}[!t]
\centering
\begin{tabular}{c||p{8.2cm}||c}
\textbf{Parameter} & \textbf{Description} & \textbf{Planck alone}\\
\hline
\hline
$\Omega_{\mathrm{b}} h^2$ & Baryon density parameter & 0.022383  \\
$\Omega_{\mathrm{c}} h^2$ & Cold dark matter density parameter & 0.12011  \\
$100 \theta_{\mathrm{MC}}$ & Angular size of the sound horizon & 1.040909 \\
$\tau$ & Optical depth to reionization & 0.0543\\
$\ln \left(10^{10} A_{\mathrm{s}}\right)$ & Amplitude of the primordial scalar perturbations & 3.0448 \\
$n_{\mathrm{s}}$ & Scalar spectral index & 0.96605 \\
\hline
$H_0\left[\mathrm{~km} \mathrm{~s}^{-1} \mathrm{Mpc}^{-1}\right]$ & Hubble constant & 67.32 \\
$\Omega_{\Lambda}$ & Dark energy density parameter & 0.6842 \\
$\Omega_{\mathrm{m}}$ & Matter density parameter & 0.3158 \\
$\sigma_8$ & Matter fluctuation amplitude on 8 $h^{-1}$ Mpc scale & 0.8120 \\
\hline
\end{tabular}
\caption{The $\Lambda$CDM best fit 6 cosmological parameters from \textit{Planck} CMB temperature and polarization power spectra. The table is divided in the main sampled and derived parameters, like $H_0$ or $\sigma_8$.}
\label{tab:cosmo_param}
\end{table}

Despite its success in explaining observational phenomena, the $\Lambda$CDM model has faced some challenges and is currently the focus of intense research. There is a lack of understanding of its two main components (Dark Matter and Dark Energy) and also, the precision level we have reached in parameter extraction, could imply potential tensions in our current understanding of the Universe \cite{giare2023cmb, hu2023hubble, abdalla2022cosmology}, which include the cosmic dipoles problem \cite{krishnan2023dipole}, the age of the Universe \cite{verde2013planck}, the lithium problem \cite{fields2011primordial}, etc\footnote{For an extensive review, see \cite{perivolaropoulos2021hubble}.}. Perhaps the most debated problem is the so-called Hubble \textit{tension}, which arises during the first release of cosmological parameters from the Planck Collaboration \cite{collaboration2014planck}, from the discrepancy of the Hubble constant found by CMB early Universe observations: $H_0=67.4 \pm 0.5 \mathrm{~km} \mathrm{~s}^{-1} \mathrm{Mpc}^{-1}$ and the value obtained by late-time surveys like SH0ES: $H_0=73.04 \pm 1.04 \mathrm{~km} \mathrm{~s}^{-1} \mathrm{Mpc}^{-1}$ \cite{riess2022comprehensive, lemos2023cosmic}. 

Understanding the late-time Universe's acceleration and its relation with the Hubble tension is a current hot discussion topic in cosmology \cite{dark2016dark, dark2005dark}. The discrepancies on its value may arise from observational systematic effects, statistical fluctuations, or new physics beyond the $\Lambda$CDM model. This is one of the reasons that, despite the $\Lambda$CDM model being a very good statistical fit that explains most of the observational phenomena, there is still interest in discovering few extension models that can explain the unresolved cosmological late-time accelerated expansion but yet recover the $\Lambda$CDM expansion history and pass the Solar System tests \cite{hwang2001f, bessa2022observational}. Many viable Modified Gravity (MG) theories feature ``screening mechanisms´´ that cause deviations from General Relativity (GR) to switch off on small scales, leaving them with significantly different predictions from GR only over cosmological distances \cite{aviles2019screenings}.

Within the MG framework, the $f(R)$ family of theories is an interesting alternative beyond $\Lambda$CDM, since it is the most natural extension performed in the Einstein-Hilbert action, at the level of the Ricci scalar, i.e. $R\rightarrow f(R)\simeq R + m R^2$ \cite{starobinsky1980new}. Some studies on cosmological constraints on these theories can be found in \cite{basilakos2013observational, jana2019constraints, nunes2017new}. A popular viable model within these theories that is in agreement with the $\Lambda$CDM expansion history and that passes the Solar System tests is the Hu-Sawicki (HS) model \cite{hu2007models}. In this case, the corresponding expansion of $f(R)$  takes the form,
\begin{equation}
f(R)=-6 \Omega_{\mathrm{DE}, 0} \frac{H_0^2}{c^2}+\left|f_{R 0}\right| \frac{\bar{R}_0^2}{R}+\ldots,
\label{eq:f(R)}
\end{equation}
where $f_{R 0}=\mathrm{d} f(R) /\left.\mathrm{d} R\right|_{z=0}$ and for values of  $\left|f_{R 0}\right| \ll 1$ the $\Lambda$CDM expansion history is well recovered. In this work, we study the performance of our architecture using simulated data for some values of $f_{R 0}$ around $\left|f_{R 0}\right|=10^{-6}$, which has proven to be still in agreement with current surveys (see, for example \cite{casas2023euclid}). An interesting point is that we can recover the $\Lambda$CDM cosmology with $|f_{R0}|=0$, then we are still in agreement with observations while we perform a comparison of models \cite{ocampo2024enhancing}.

Furthermore, the $\Lambda$CDM model also fails to give an explanation about the nature of the origin of the Universe. The anisotropies of the CMB indicates that the primordial density perturbations that populated the very early Universe should have been almost Gaussianly distributed, explaining the observed homogeneity and isotropy of the cosmos. Inflation is widely regarded as the mechanism responsible for the origin of these primordial density perturbations and, by construction, solves the horizon and flatness problems. This exponentially rapid expansion phase took place fractions of a second after the Big Bang provides a solid framework for the generation of primordial density fluctuations that raised the current LSS. These fluctuations are imprinted on the primordial power spectrum (PPS), which describes the distribution of these density perturbations across different scales. In the simplest inflationary model, vanilla canonical single-field inflation, this power spectrum is usually defined as:
\begin{equation}
P_{\mathcal{R}, 0}(k)=\frac{2 \pi^2}{k^3} \mathcal{P}_{\mathcal{R}, 0}(k)= A_{\mathrm{s}}\left(\frac{k}{k_{\star}}\right)^{n_{\mathrm{s}}-1},
\label{eq:PPs}
\end{equation}
where $n_s$ and $A_s$ are the spectral index and the amplitude of the comoving curvature perturbations, respectively, and $k_{\star} = 0.05\, \mathrm{Mpc}^{-1}$ is the pivot scale.

In this sense, \autoref{eq:PPs} is compatible with the assumptions of the $\Lambda$CDM, setting a phenomenological parametrisation of an \textit{almost}-scale invariant power spectrum. However, it lacks an intrinsic mechanism to explore further inflationary scenarios or further alternatives, making the study of deviations from this power-law a hot topic in cosmology \cite{features_review}.

Over the past years, an explosion of data has significantly contributed to numerous studies for testing deviations from $\Lambda$CDM. These works have been carried out with data from late time Universe, like BOSS \cite{alam2017clustering}, DES \cite{abbott2022dark}, KiDS \cite{de2013kilo} and early Universe with the characterization of CMB flcutuations with WMAP \cite{hinshaw2013nine}, Planck \cite{aghanim2020planck} and ACT \cite{aiola2020atacama}. In the future, it is expected that the most precise constraints of the cosmological model parameters will come from stage IV surveys such as the ESA \textit{Euclid} mission \cite{blanchard2020euclid, laureijs2011euclid, mellier2024euclid}, LSST \cite{ivezic2019lsst} and DESI \cite{abareshi2022overview}, which are promising to constrain extensions beyond the $\Lambda$CDM model. 

This explosion of data collection, beyond allowing an increase in quantity and quality, will pose questions on how to efficiently analyse the data and how to adapt computational pipelines. This is one of the main reasons why Machine Learning (ML) has become more popular in cosmology in the past years. The primary benefits of ML techniques lay in its ability to uncover patterns, insights, and knowledge from large datasets without explicit instructions, and to adapt and enhance its performance independently. Therefore, ML methods can fully unlock the potential of multi-probe cosmology by effectively improving systematic effects and optimally integrating information from various surveys \cite{joseph2021joint}. It facilitates the detection and classification of cosmological sources as well as the extraction of information from images \cite{dvorkin2022machine, moriwaki2023machine}. ML methods have also been introduced to cosmological simulations for robust predictions (see \cite{rose2024introducing}). Additionally, some work has been done in the context of cosmological parameter inference \cite{murakami2023non, min2024deep, garcia2024bayesian} and for testing models beyond $\Lambda$CDM in the context of galaxy clustering \cite{ocampo2024enhancing} and weak lensing \cite{peel2019distinguishing}. Furthermore, some studies using CMB data and ML were performed, for example for lensing reconstruction \cite{yan2023lensing}, also for the reconstruction of the inflationary potential using genetic algorithms \cite{kamerkar2023machine}, for the removal of the CMB foreground components \cite{remazeilles2011foreground,wang2022recovering,yan2023recovering}, for CMB data selection coming from individual detector time-streams \cite{rojas2020classifying}, CMB component separation \cite{mccarthy2024signal}, autoencoders to extract information about the cosmological parameters from the power spectrum \cite{de2022cosmic}, and deep learning of CMB radiation temperature \cite{salti2022deep}. While other studies of probing primordial features with future galaxy surveys were done in \cite{ballardini2016probing, ballardini2018probing}, this is the first time a study of this kind has been performed to test beyond $\Lambda$CDM models using machine learning on the CMB angular power spectra.

The importance of this work relies on testing models beyond-$\Lambda$CDM using a classification approach based on Machine Learning Neural Networks, with interpretability techniques. This approach can be an important step forward for efficiently exploring theoretical extensions to the standard model without immediately relying on the computational demands of full Bayesian statistical analyses to sample the posterior distribution of the parameters of interest. Bayesian methods provide rigorous tools for parameter estimation and model comparison, but they require significant computational resources and often assume that the data is sufficiently sensitive to distinguish between competing models. By employing both model classification and interpretability as a precursor, we can identify the data most sensitive to the models we aim to test.

While the official operations of the ESA Planck mission came to an end, after showing the capacity of the CMB temperature and polarisation anisotropies in constraining the underlying cosmological model, it can still be leveraged to provide stringent tests for theories of fundamental physics beyond the $\Lambda$CDM. In the context of the late-time Universe we can test MG models related to dark energy, like the Hu-Sawicki model, and in the case of the early Universe we can use the data to search for deviations in the primordial power spectrum power-law (the so-called features). Therefore, the aim of this work is to test both cases using Planck 18 CMB angular power spectrum, to perform ML- based model selection. For this, we train and test a Neural Network with first, HS simulated data vs $\Lambda$CDM simulated data and second, feature models simulated data vs $\Lambda$CDM, both cases taking into account the Planck 18 data uncertainties.

The layout of our paper is the following. In \autoref{sec:methodology}, we introduce the two beyond-$\Lambda$CDM cases we want to test, the Hu-Sawicki model and the linear feature model in the primordial power spectrum (\autoref{subsec:deviations}), and we explain how we simulated our training data and created our neural network architecture (\autoref{subsec:MLpipeline}). In \autoref{sec:results}, we show the results of the analysis and comment on the main characteristics found: in \autoref{subsec:HSmodel} and \autoref{subsec:Featuremodel} we discuss the results from the analysis of the Hu-Sawicki model and the primordial feature template respectively, and, finally, we discuss the interpretability of our machine learning pipeline in \autoref{subsec:Interpretability}. Finally, in \autoref{sec:conclusions}, we summarise our findings, explain the main lessons learnt and draw possible future work to further exploit our Machine Learning pipeline.

\begin{figure}[H]
    \centering
    \subfloat{\includegraphics[width=0.81\linewidth]{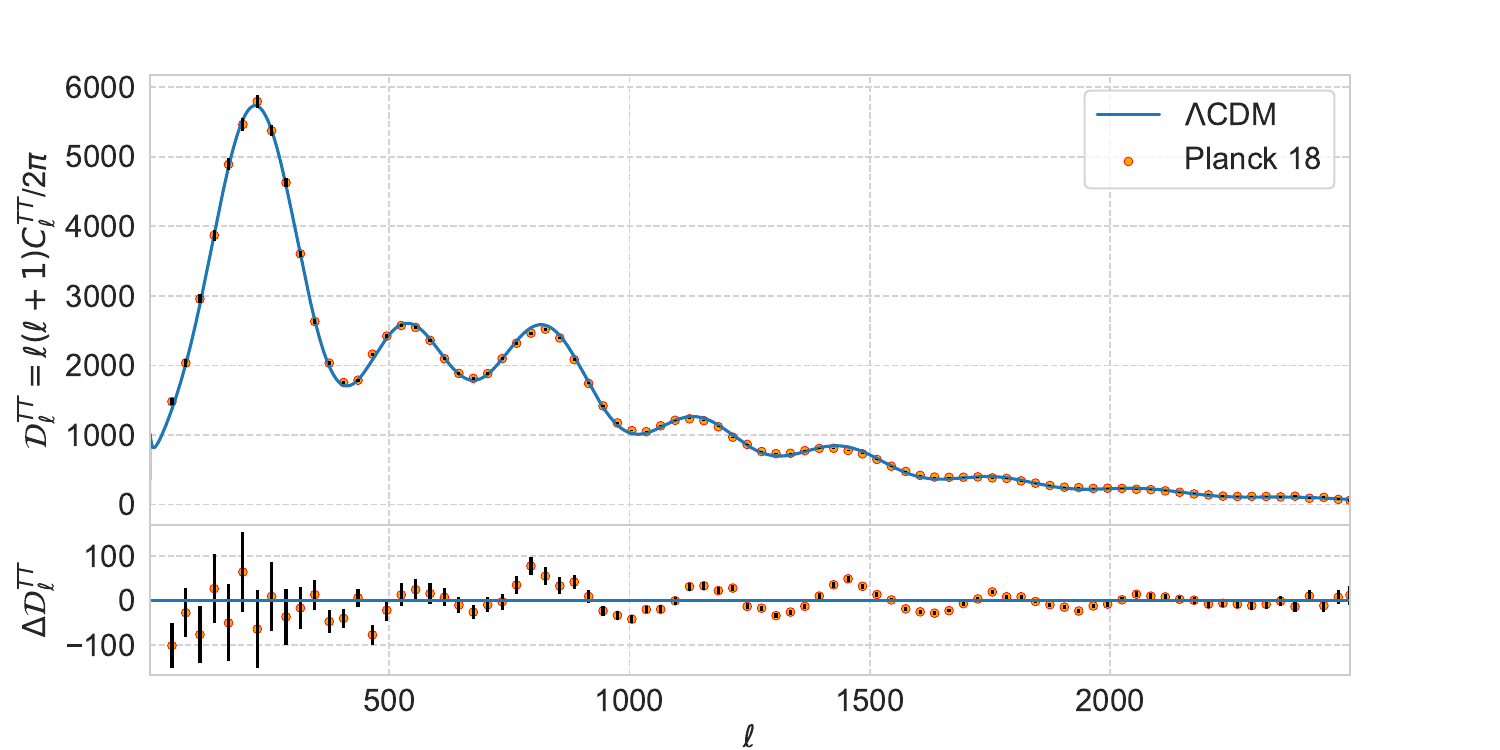}}\\
        \label{fig:Planck_TT}
    \centering
    \subfloat{\includegraphics[width=0.81\linewidth]{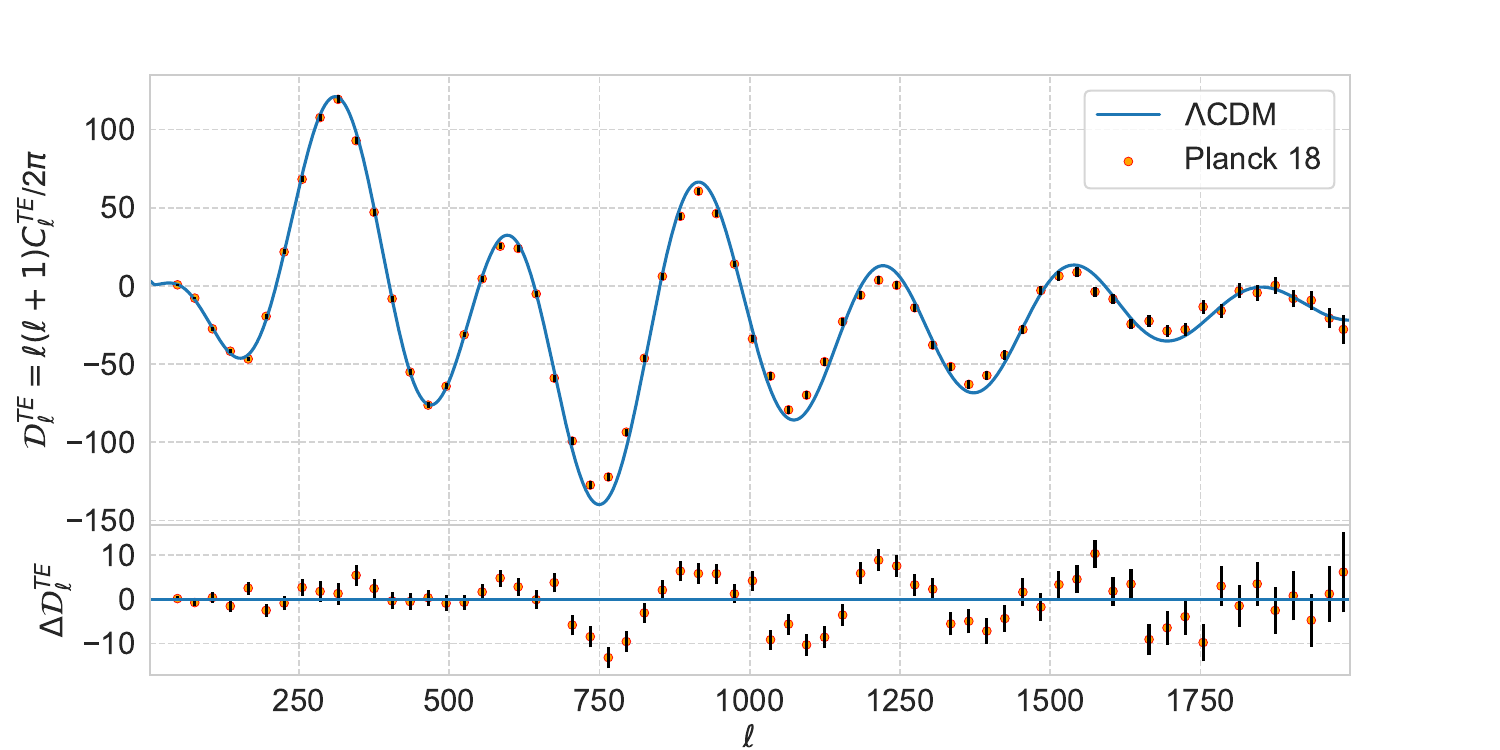}}\\
    \label{fig:Planck_TE}
    \subfloat{\includegraphics[width=0.81\linewidth]{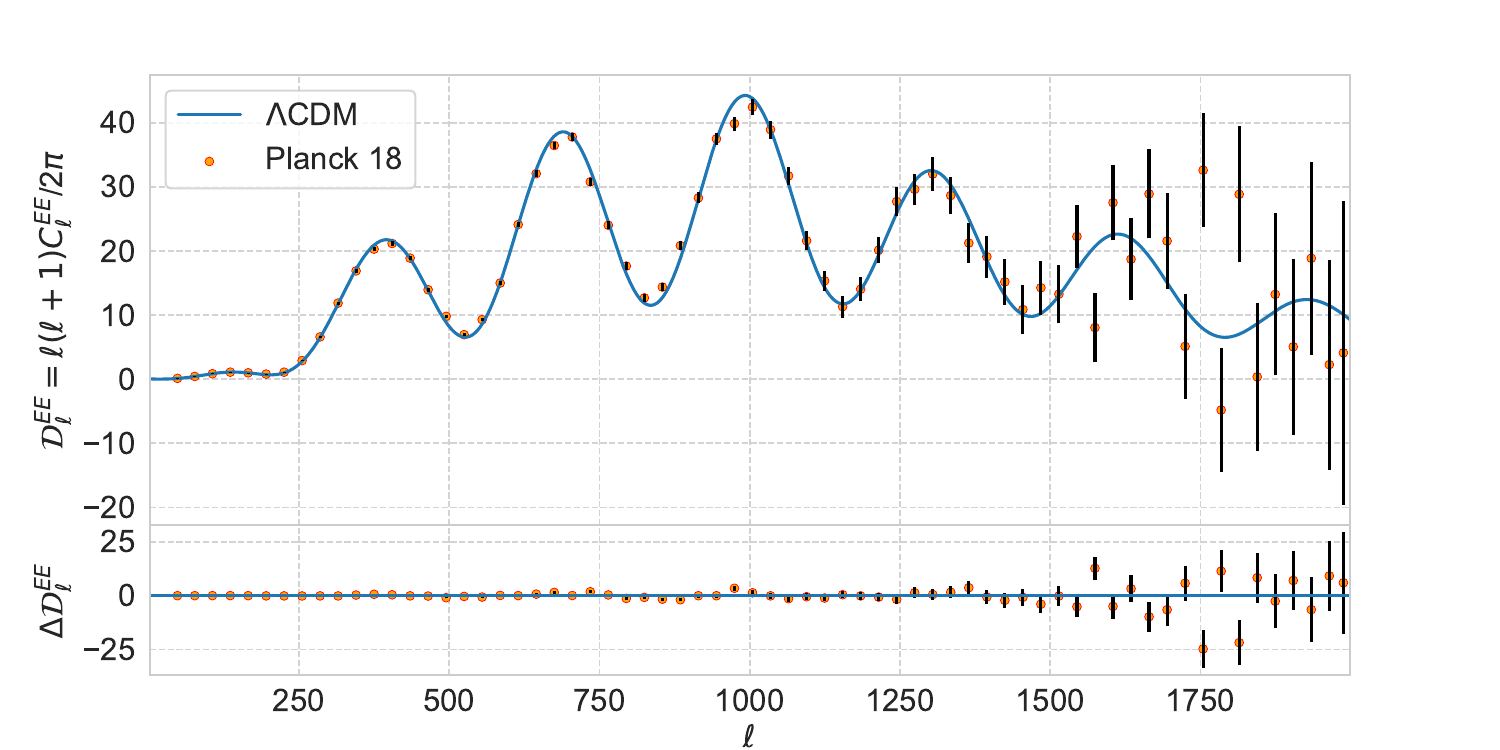}}
    \label{fig:Planck_EE}
    \caption{Top panels: CMB angular power spectra Planck 18 binned temperature TT, polarisation EE, and cross-correlations TE anisotropies (orange dots, with error bars), jointly with the best theory prediction from $\Lambda$CDM (solid blue line), calculated with \texttt{CLASS}. Bottom panels: difference with respect to the best fit for all the different components.}
    \label{fig:CLASS_Planck_TT_TE_EE}
    \label{fig:TT_TE_EE}
\end{figure}

\section{Methodology}\label{sec:methodology}
The focus of this study is the development and performance evaluation of a robust data analysis pipeline for machine learning-based model selection, in order to study two cases, (1) deviations from general relativity within the framework of the Hu-Sawicki model and (2) searching for primordial features in the primordial power spectrum. To achieve this, we implemented an architecture based on Neural Networks to work with Planck 18 data at the level of the CMB angular power spectra $C_{\ell}$. The angular power spectra is given by \cite{baumann2018tasi, mactavish2006cmb}
\begin{equation}
C_{\ell}^{X Y}=\frac{2}{\pi} \int k^2 \mathrm{~d} k \,P_{\mathcal{R}}(k) \Delta^{X}_{\ell}(k) \Delta^{Y}_{\ell}(k),
\label{eq:C_ells}
\end{equation}
where XY stands for the TT, TE and EE polarisation patterns, $\,P_{\mathcal{R}}(k)$ is the power-law primordial spectrum given by \autoref{eq:PPs}, and $\Delta^{X}_{\ell}(k)$ is the transfer function that is related to the evolution of primordial density fluctuations from their primordial state to the time of recombination and beyond, and $\ell$ is the multipole moment \cite{baumann2009tasi, mactavish2006cmb}. We can also rewrite the angular power spectra in the common $\mathcal{D}^{XY}$ definition:
\begin{equation}\label{eq:dells}
\mathcal{D}^{XY}_{\ell}=\frac{\ell(\ell+1)C_{\ell}^{XY}}{2\pi}.
\end{equation}

The transfer function $\Delta_{X \ell}(k)$ is usually calculated using so-called Boltzmann solvers \texttt{CAMB} \cite{camb} or \texttt{CLASS} \cite{class}. An important aspect of \autoref{eq:C_ells} is that any late-time Universe or modified gravity model is reflected in the transfer function, subsequently affecting the angular power spectra, whereas any deviation originating from the early Universe physics impacts the power-law primordial power spectrum $P_{\mathcal{R}}(k)$, which also induces changes in the angular power spectra. This interplay makes it an ideal framework for testing both scenarios.

We are interested in studying the statistical properties of the CMB anisotropies, characterized by the temperature (TT mode), polarization (EE mode) and cross (TE) angular power spectra, since the position and amplitude of the peaks and dips of these spectra are sensitive to the assumed underlying cosmological model. See, for example, \autoref{fig:TT_TE_EE}, which shows the Planck 18 angular power spectrum binned data\footnote{The data can be found at the Planck Legacy Archive at \url{http://pla.esac.esa.int/pla}.} compared to the best-fit prediction of the $\Lambda$CDM model. For every computation of the $\Lambda$CDM baseline model, we chose the fiducial cosmological parameters according to Planck 18 cosmological results \cite{aghanim2020planck}, as noted in \autoref{tab:cosmo_param}, and the software \texttt{CLASS}.

\subsection{Theoretical predictions in beyond-$\Lambda$CDM models}\label{subsec:deviations}
\subsubsection{Deviations from general relativity}
The scenario to test for deviations from GR that we will stick to is the HS $f(R)$ model, having an effect on the transfer function that at the same time has an impact on the angular power spectrum \autoref{eq:C_ells} \cite{kou2024constraining}. To see this, we consider a modification to the Einstein-Hilbert action of the form \cite{hu2007models}: 
\begin{equation}
S=\int \mathrm{d}^4 x \sqrt{-g}\left[\frac{1}{2 \kappa}f(R)+\mathcal{L}_{\mathrm{m}}\right],
\end{equation}
where $R$ is the Ricci scalar, $\mathcal{L}_{\mathrm{m}}$ is the matter Lagrangian and $\kappa=8\pi G_N$ is a constant (with $G_N$ being the Newton's constant). By varying the action and taking some considerations, like the sub-horizon and quasi-static approximations, we get to the following second-order differential equation \cite{tsujikawa2007matter}:
\begin{equation}
\ddot{\delta}+2 H \dot{\delta}-4 \pi G_{\text {eff }} \rho \delta=0, \\
\end{equation}
where the dot represents the derivative with
respect to the cosmic time $t$ and,
\begin{equation}
G_{\text {eff }}=\frac{G_N}{F} \frac{1+4 \frac{k^2}{a^2} \frac{F^{\prime}}{F}}{1+3 \frac{k^2}{a^2} \frac{F^{\prime}}{F}}.
\end{equation}
is the effective gravitational constant that accounts for the effect of modified gravity \cite{motohashi2011f,tsujikawa2007matter}. Here $F(R) \equiv f^{\prime}(R)$, the prime denotes derivative with respect to $R$, and $a$ is the scale factor. Therefore, in this context, the only part affected by $f(R)$ is the transfer function. See \cite{song2007large} for a reference and \cite{ravi2024investigating,kumar2023new,wang2022pantheon+} for current constraints on this model.

\subsubsection{Search for primordial features}
\begin{figure}[h!]
    \centering \includegraphics[width=0.9\linewidth]{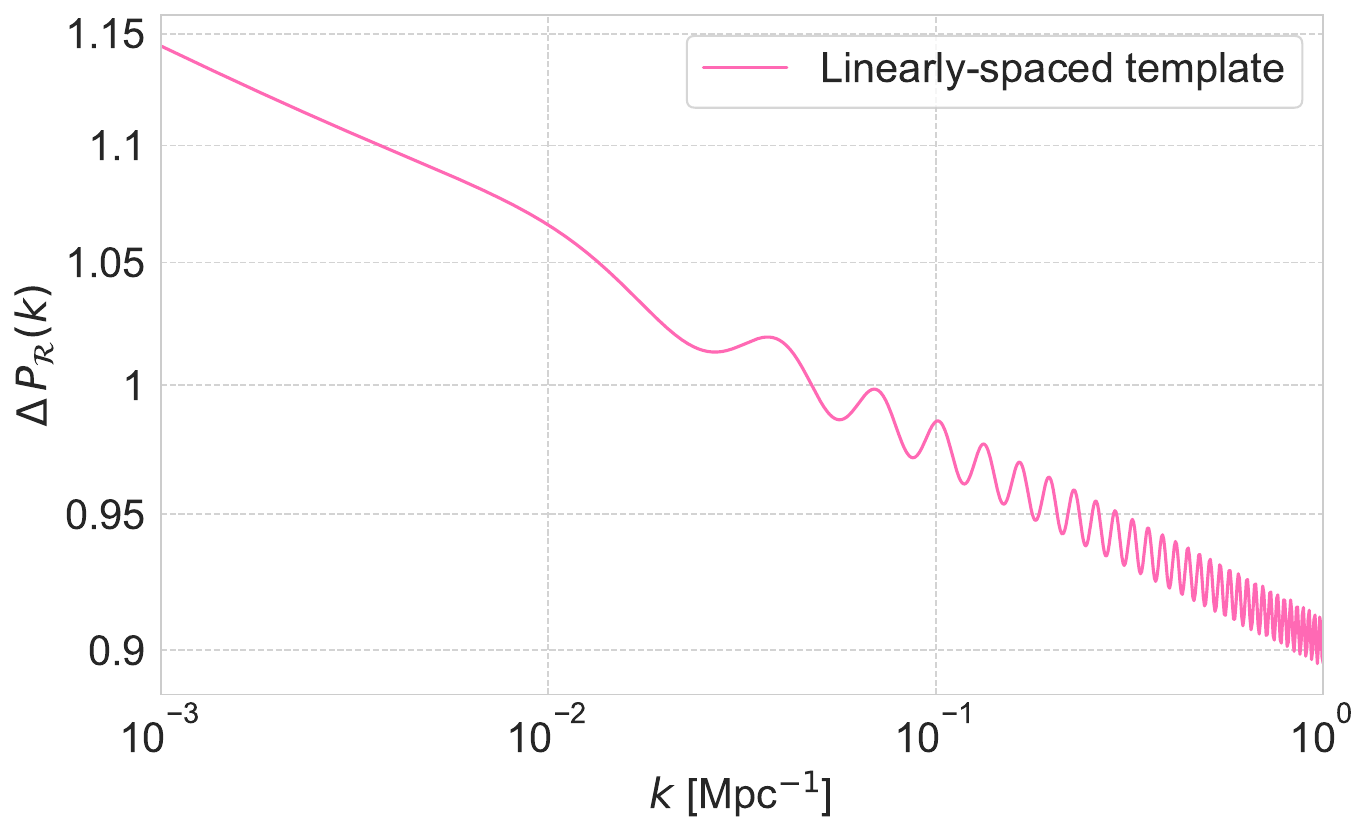}
    \caption{Linearly-spaced primordial feature $\Delta P_\mathcal{R}$ on the power spectrum of curvature perturbations, given the fiducial values at \autoref{eq:lin_oscillations}, as a function of scale, in logarithmic space.}
    \label{fig:Feature}
\end{figure}

All cosmological observations so far align with the assumption of adiabatic, Gaussian, and nearly scale-invariant initial conditions. These results strongly support a specific inflationary scenario known as slow-roll inflation. This pattern is broadly accepted as matching the predictions of the simplest inflationary single-field canonical model \cite{lodha2023searching, kamerkar2023machine}. However, since our observations only capture perturbations, deducing the background that produced them reveals that various underlying theories can lead to the same set of cosmological observables. If there are any features in the primordial power spectrum (deviations from the nearly scale-invariant power spectrum), they would provide a distinctive and revealing insight into the fundamental theory behind the mechanism that created these initial fluctuations \cite{features_review}. Primordial features, $\Delta P_{\mathcal{R}}/P_{\mathcal{R}, 0}$, are parametrised as small deviations from the power-law primordial power spectrum introduced in \autoref{eq:PPs} as:
\begin{equation}
P_{\mathcal{R}}(k)=P_{\mathcal{R}, 0}(k)\left[1+ \frac{\Delta P_{\mathcal{R}}}{P_{\mathcal{R}, 0}}\left(k\right)\right],
\label{eq:PPs_feature}
\end{equation}
where $P_{\mathcal{R}, 0}$ is given by \autoref{eq:PPs_feature}. There are several theories that predict different shapes for features (localised, linear or logarithmic spaced ones). In this paper, we study a toy-model feature template with oscillations linearly spaced in Fourier space with a constant amplitude, superimposed on the power-law (see \autoref{fig:Feature}):
\begin{equation}
\frac{\Delta P_{\mathcal{R}}}{P_{\mathcal{R}, 0}}=A_{\operatorname{lin}} \sin \left(\omega_{\operatorname{lin}} \frac{k}{k_{\star}} +\phi\right),
\label{eq:Feature}
\end{equation}
where $A_{\operatorname{lin}}$ is the amplitude of the feature, $\phi$ is some arbitrary phase and $\omega_{\operatorname{lin}}$ is the frequency of oscillations. The fiducial values of the parameters that define the linear oscillations $\Theta_{\text {lin }}$ are given by \cite{ballardini2024euclid},
\begin{equation}
\Theta_{\text {lin }}=\left\{{A}_{\text {lin }}=0.01, \omega_{\text {lin }}=10, \phi_{\operatorname{lin}}=0\right\} .
\label{eq:lin_oscillations}
\end{equation}

In this work, we also study the performance of our NNs using simulated data for some values of ${A}_{\text {lin }}$ around the fiducial, which is also in agreement with current surveys \cite{ballardini2024euclid}. Similar to the Hu-Sawicki case, we can also recover the $\Lambda$CDM cosmology with ${A}_{\text {lin }}=0$.

The discovery of any additional inflationary signals hidden in the data could significantly change our understanding of the early stages of the Universe. While, several feature templates have been tested against Planck 18 data \cite{akrami2020planck}, developing an alternative pipeline to test features could offer a way to test extensive photometric and spectroscopic surveys, which offer a complementary perspective on large-scale structures (LSS). This enhances the sensitivity to high-frequency signals and complements CMB measurements.

\subsection{Machine Learning pipeline using Neural Networks}\label{subsec:MLpipeline}
The Planck mission provided detailed measurements of the anisotropies of the CMB, allowing to perform cosmological inference on dark energy and modified gravity \cite{ade2016planck} and inflation \cite{collaboration2014planck} models, as well as putting the tightest constraints on cosmological parameters \cite{aghanim2020planck} by sampling the posterior distribution of the parameters of interest using Markov Chain Monte Carlo (MCMC) exploration of the likelihood. Instead, this work has the objective to test GR deviations from $\Lambda$CDM and search for primordial features leveraging the power of Neural Networks (NN). NNs are known to be a powerful tool consisting of many connected units known as neurons that produce a sequence of real-valued activations, and are widely used in classification tasks. Within these algorithms, deep neural networks are one of the most popular and are characterized for having fully connected (dense) layers, i.e. all possible connections layer-to-layer are present \cite{schmidhuber2015deep}. Deep NN can enhance the accuracy of model predictions and improve our understanding of the physical processes behind certain patterns. This is an important task in cosmology, since distinguishing between models could potentially lead to some insights into the nature of the evolution of the Universe, dark matter and dark energy, and could set a preliminary starting point for an exhaustive (classical) statistical analysis on a later time. The analysis pipeline and datasets are publicly available on GitHub\footnote{\url{https://github.com/IndiraOcampo/CMB_ML_based_model_selection.git}} and Zenodo\footnote{Primordial feature template (\url{10.5281/zenodo.13829664}), and Hu-Sawicki (\url{10.5281/zenodo.13828966})}, respectively.

\subsubsection{Simulated datasets}\label{subsub:simulateddata}
To train and test our NN, we simulated two types of datasets:\footnote{Note that we use the term \textit{datasets} to refer to the synthetic self-generated data used to train the neural network (using Planck 18 uncertainties), and it should not be mistaken with the real Planck 18 data.}

\begin{itemize}
    \item \textbf{Deviations from general relativity (Hu-Sawicki model):} We generated the dataset using \texttt{MGCLASS}\footnote{\url{https://gitlab.com/zizgitlab/mgclass--ii/s}} \cite{sakr2022cosmological}, a modified version of the Boltzmann solver \texttt{CLASS}, within the framework of the HS model. We simulated 1000 samples (since the scalability of the code does not compensate the performance of the NN) for $\Lambda$CDM baseline model (see \autoref{tab:cosmo_param}) varying $\Omega_{\text{cdm}}$ within the range $[0.1,0.15]$, and 1000 for the $f(R)$ model with grids of values $|f_{R0}|\in[10^{-6},5\times 10^{-6}]$, for three components $C_{\ell}^{TT}$, $C_{\ell}^{TE}$ and $C_{\ell}^{EE}$. For this case, Planck 18 uncertainties derived from the corresponding covariance matrix, are subsequently incorporated into the data. Finally to study the point in which the NN performance breaks down for this section, we also generated data for $|f_{R0}|$ for different ranges spanning $[10^{-4},5 \times 10^{-4}]$, $[10^{-6},5 \times 10^{-6}]$, $[10^{-7},5 \times 10^{-7}]$ and $[10^{-8},5 \times 10^{-8}]$.
    \item \textbf{Search for primordial features (linearly-spaced template):}
    We generated a synthetic dataset using a modified version of the Boltzmann solver code \texttt{CLASS} \cite{class} by introducing the target linearly-spaced feature in the primordial power spectrum (\autoref{eq:Feature}). We used the Planck best fit cosmological parameters \autoref{tab:cosmo_param} to generate 500 samples (once again, because the scalability of the code does not compensate the performance of the NN), of the angular power spectrum for $\Lambda$CDM varying $\Omega_{\text{cdm}}\in[0.05,0.15]$, and 500 for the linearly-spaced primordial feature, varying $\Omega_{\text{cdm}}$ in the same range and $A_{\operatorname{lin}} \in [10^{-2},5 \times 10^{-2}]$. We followed this procedure for each of the three components: $C_{\ell}^{TT}$, $C_{\ell}^{TE}$ and $C_{\ell}^{EE}$ and added the uncertainties from the covariance matrix set out of the Planck errors. We also aimed to test the point in which the NN performance breaks down, so we also generated data for $A_{\operatorname{lin}}$ for different ranges between $[10^{-3},5 \times 10^{-3}]$, $[10^{-4},5 \times 10^{-4}]$, $[10^{-5},5 \times 10^{-5}]$ and $[10^{-6},5 \times 10^{-6}]$.
\end{itemize}

\subsubsection{Neural Network architecture}\label{subsubsec:NNarchitecture}
\begin{figure}[h!]
    \centering\includegraphics[width=0.19\linewidth]{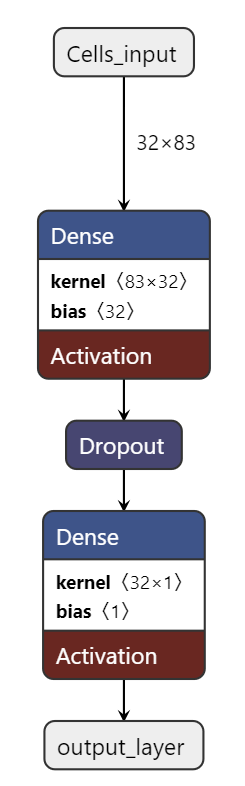}
    \caption{NN architecture for the TT case with 83 neurons in the input layer (that accounts for the 83 data points of each $C_\ell^{TT}$ realization). We set the batch size to the default 32 value, followed by the implementation of a fully connected layer of 32 neurons, a dropout layer and an output layer with one neuron.}
    \label{fig:NN_architecture}
\end{figure}

Different architectures can influence how effectively a NN can capture underlying patterns and features within the data. In this paper, we chose deep neural networks, because of their potential in classification tasks. We developed 4 different deep NN: one for each CMB angular power spectra case (TT, TE, EE), and a final one for the joint case (TT + TE + EE). A generalized scheme of the implemented NN's architecture is shown in \autoref{fig:NN_architecture}, where the units (neurons) in the input layer depended on the number of data points of each case: for $C_{\ell}^{TT} \--$ 83 units, for $C_{\ell}^{TE} \--$ 66 units, for $C_{\ell}^{EE} \--$ 66 units and the last one for the combined analysis $C_{\ell}^{TT}$ + $C_{\ell}^{TE}$ + $C_{\ell}^{EE} \--$ 215 units. The units correspond to the number of data points present in Planck 18 binned data.\\

The efficiency of training a deep NN heavily relies on the quality of the input data. Therefore, it is essential to assess, clean, and process the data to maximize the NN's ability to learn key features. Common approaches for achieving this are statistical normalization techniques. In this case, we applied the so-called Z-score normalization for the data pre-processing, which consists on measuring how many standard deviations $\sigma$ a data point is from the mean of the dataset $\mu$, and use this reference setup to effectively re-scale the entire dataset so that $\mu=0$ and $\sigma=1$ \cite{patro2015normalization}. 

\begin{figure}[h!]
\centering
\subfloat{\includegraphics[width=0.52\linewidth]{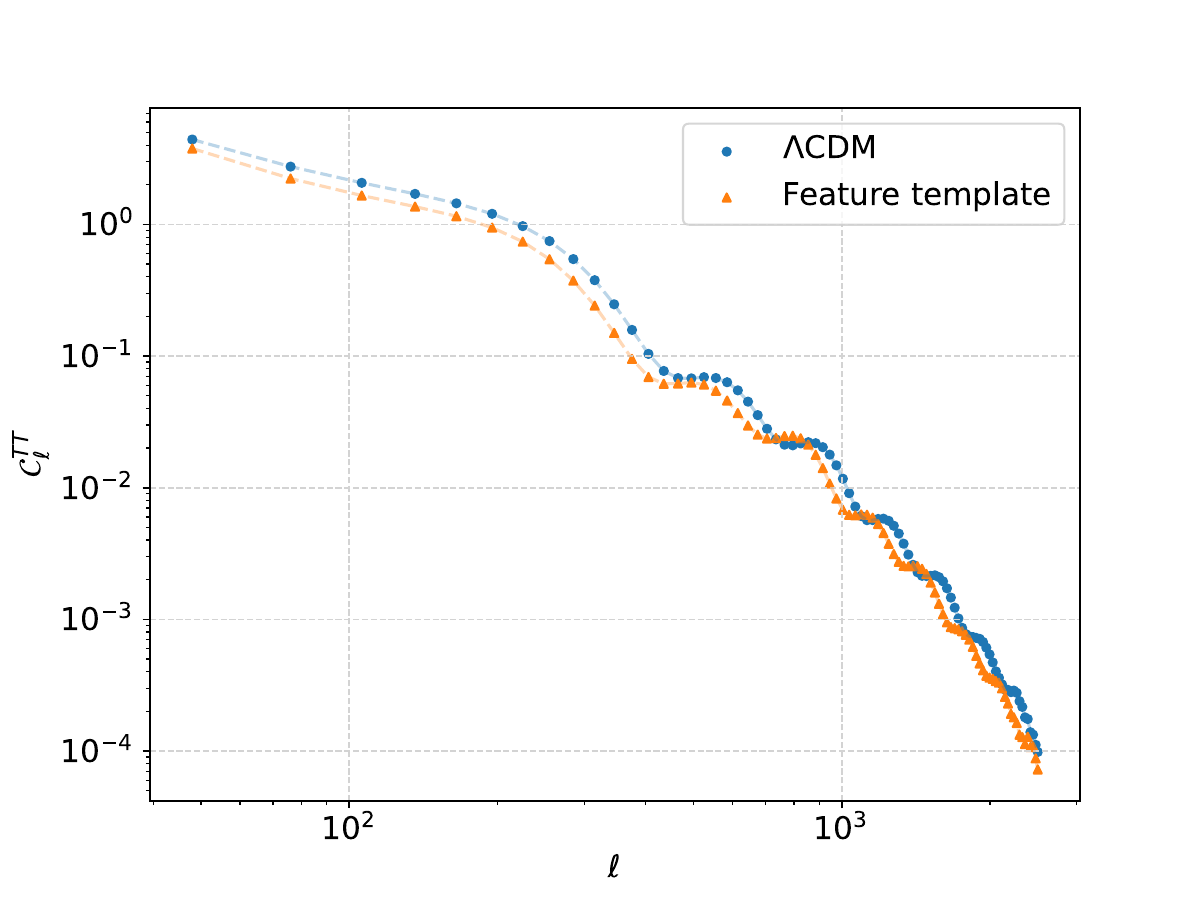}}
\subfloat{\includegraphics[width=0.52\linewidth]{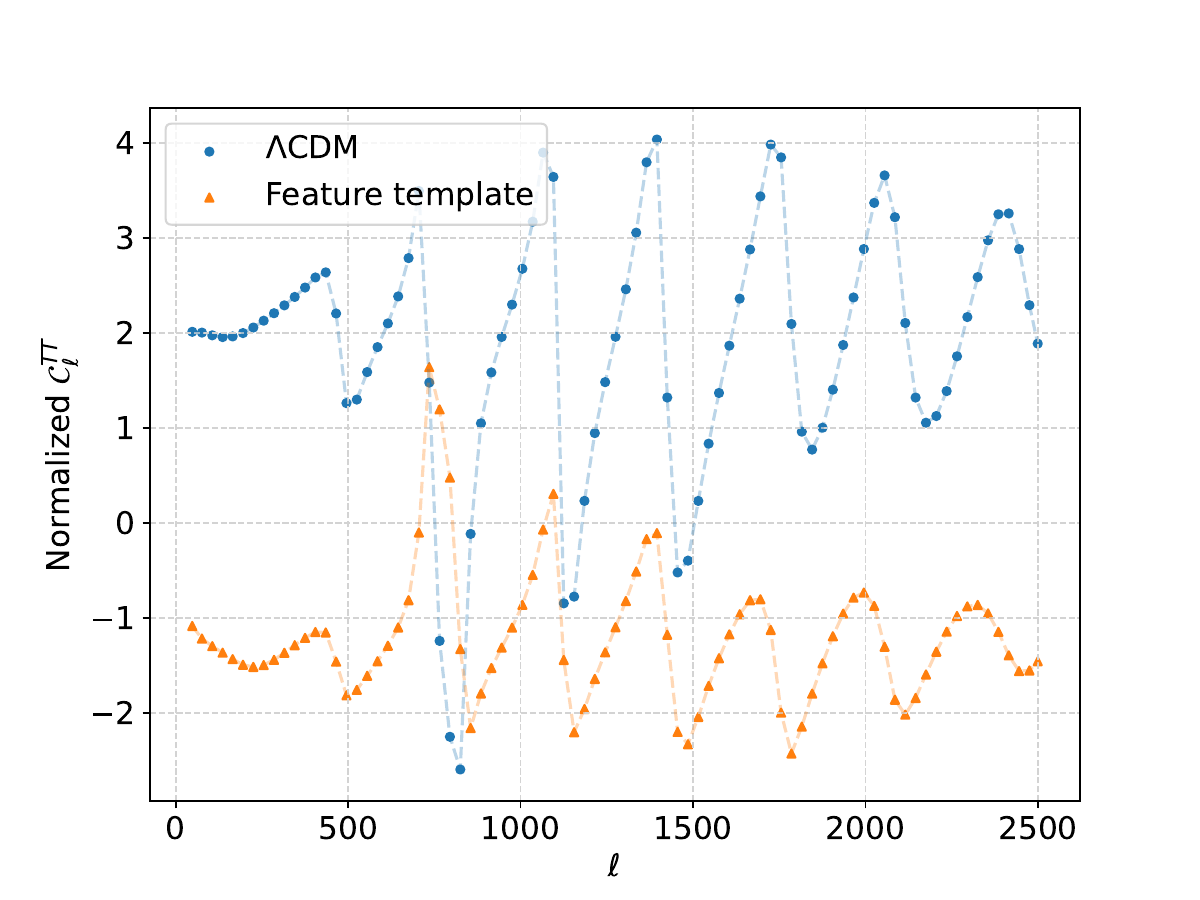}}
\caption{Illustration of how the Z-score normalization for the $C_{\ell}^{TT}$ CMB power spectrum maximize the differences between $\Lambda$CDM and extension models (in this case, the linearly-spaced feature template as in \autoref{eq:Feature}). Right panel: $C_{\ell}^{TT}$ wihtout normalization(in logarithmic space). Left panel: normalized $C_{\ell}^{TT}$.} The Z-score normalization not only re-scales the mean of the $C_{\ell}^{TT}$ but also highlights the difference in phase for the $C_{\ell}^{TT}$ values as a function of $\ell$.
\label{fig:normalization.Z-score}
\end{figure}

The normalization with a global reference for $\mu$ and $\sigma$ guarantees that data points across all redshift bins are placed on the same footing, ensuring comparability across the entire dataset. To exemplify this in our analysis, we show the impact of using Z-score normalization in one of our datasets; in particular, on the $C_{\ell}^{TT}$ component for the feature model case (see \autoref{fig:normalization.Z-score}). The differences on $C_{\ell}^{TT}$ after normalization between $\Lambda$CDM and the linearly-spaced feature template are more evident, allowing the deep NN to easily identify possible features for further classification and identification. Similar results are found for the HS model. We performed the same Z-score normalization for the pre-processing of all 8 architectures (4 for the HS model and 4 for the primordial feature model). Furthermore, we have assessed the impact of using the Z-score normalization on the $\mathcal{D}_\ell$ (see \autoref{eq:dells}) instead of $C_{\ell}$ to train the NN, not finding significant differences in the output and therefore, re-assuring the stability of the NN architecture.

After completing the data pre-processing steps, we divided the 1500 data realizations (with $50\%$ corresponding to $\Lambda$CDM and $50\%$ to either HS or primordial features, depending on the scenario) into two sets: $70\%$ for training and $30\%$ for testing. We set the batch size to 32, and incorporated a hidden dense layer with 32 units with a ReLU activation function \cite{agarap2018deep}. We included a dropout layer (with a dropout rate of 0.2) as a regularisation technique to prevent over-fitting \cite{srivastava2014dropout}. Finally, we added the last dense layer with one unit and a sigmoid activation function that enhances the classification score for the model: $0 \--$ $\Lambda$CDM and $1 \--$ HS or primordial feature. The models were compiled with an Adam optimizer ($\text{learning}\_\text{rate} = 0.0002$), a binary cross entropy loss function, and trained over 1000 epochs. See \autoref{fig:NN_architecture} for a schematic view of one of the NN's architecture.

\section{Results and discussion}\label{sec:results}
In this section, we present the main results obtained with our pipeline and the outcome of the learning assessment with ML interpretability. The classification performance of our NN's architecture is available in \autoref{fig:confusion_matrices} for the three components $C_{\ell}^{TT}$, $C_{\ell}^{TE}$, $C_{\ell}^{EE}$ and its combination $C_{\ell}^{TT} + C_{\ell}^{TE} + C_{\ell}^{EE}$, for both the two cases under study (deviations from general relativity and features in the primordial power spectrum). We used ranges of parameters that encompassed the fiducial values of both models as $A_{\operatorname{lin}}=10^{-2}$ and $|f_{R0}|=10^{-6}$ (see \autoref{fig:FR0_Feat_Cls}). From these results, we can see that the NNs perform better in detecting the alternative initial conditions in the context of the early Universe physics than the deviations from GR in the late-time Universe. 

\subsection{Deviations from general relativity: Hu-Sawicki model}\label{subsec:HSmodel}
\begin{figure}[H]
\centering
\subfloat{\includegraphics[width=0.5\linewidth]{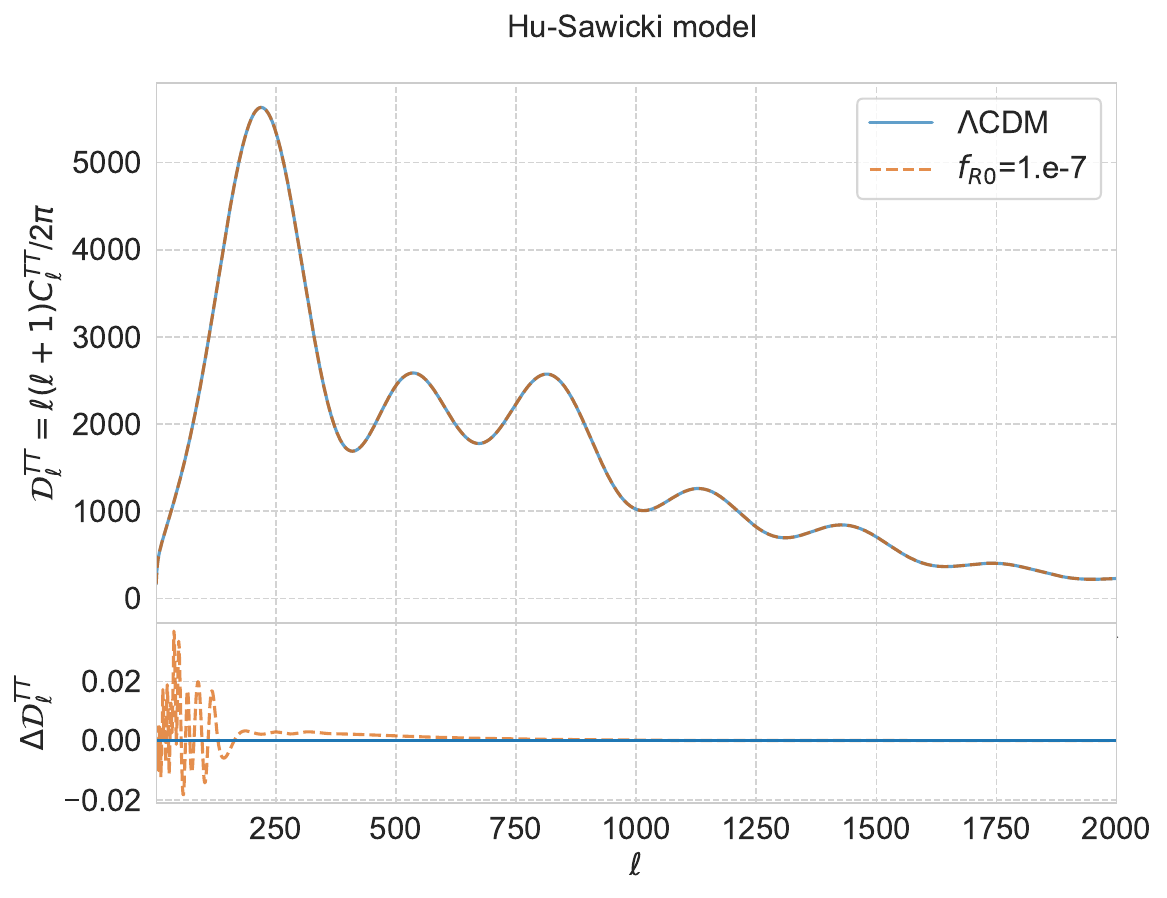}}
\subfloat{\includegraphics[width=0.525\linewidth]{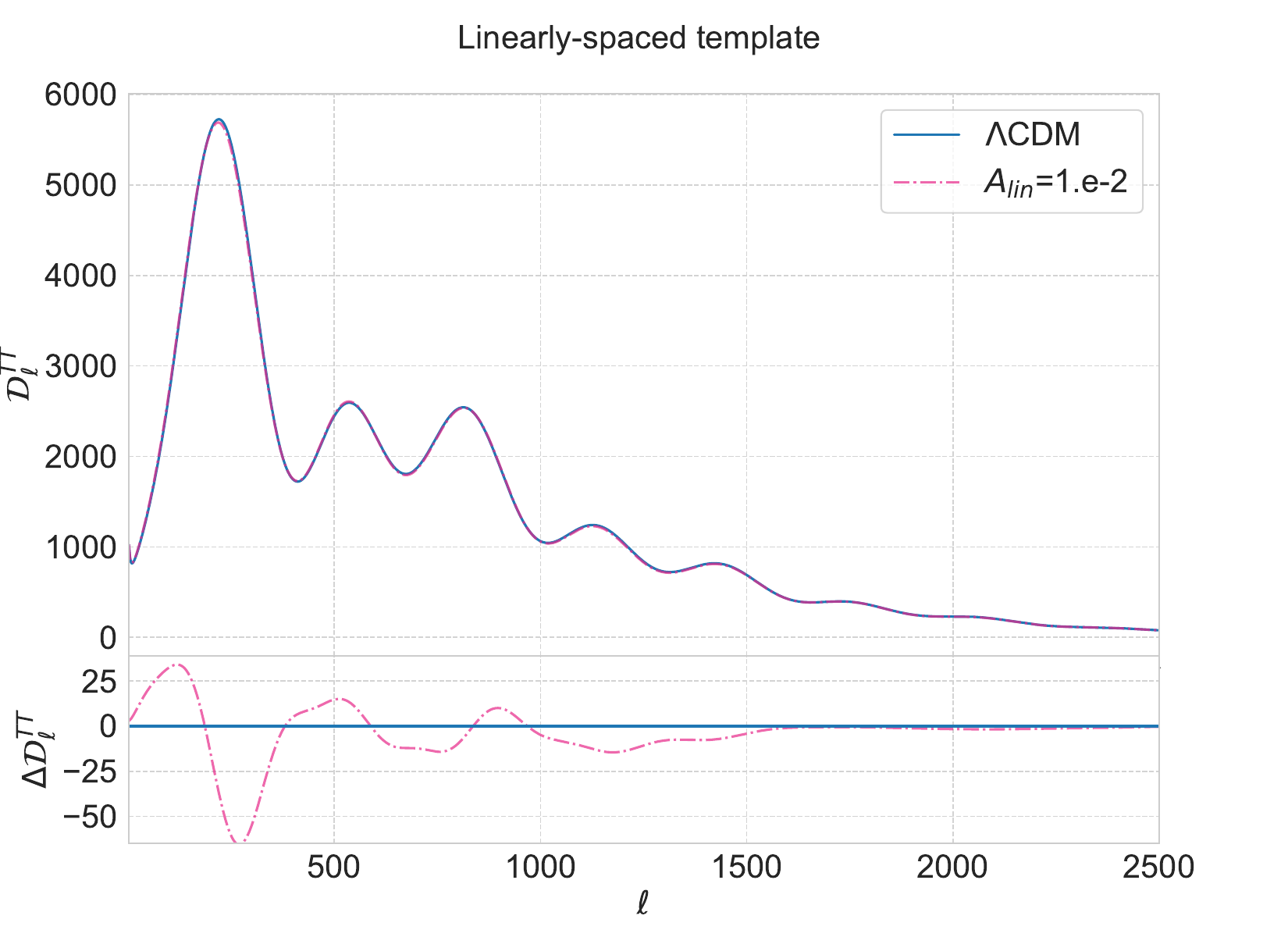}}\\
\subfloat{\includegraphics[width=0.5\linewidth]{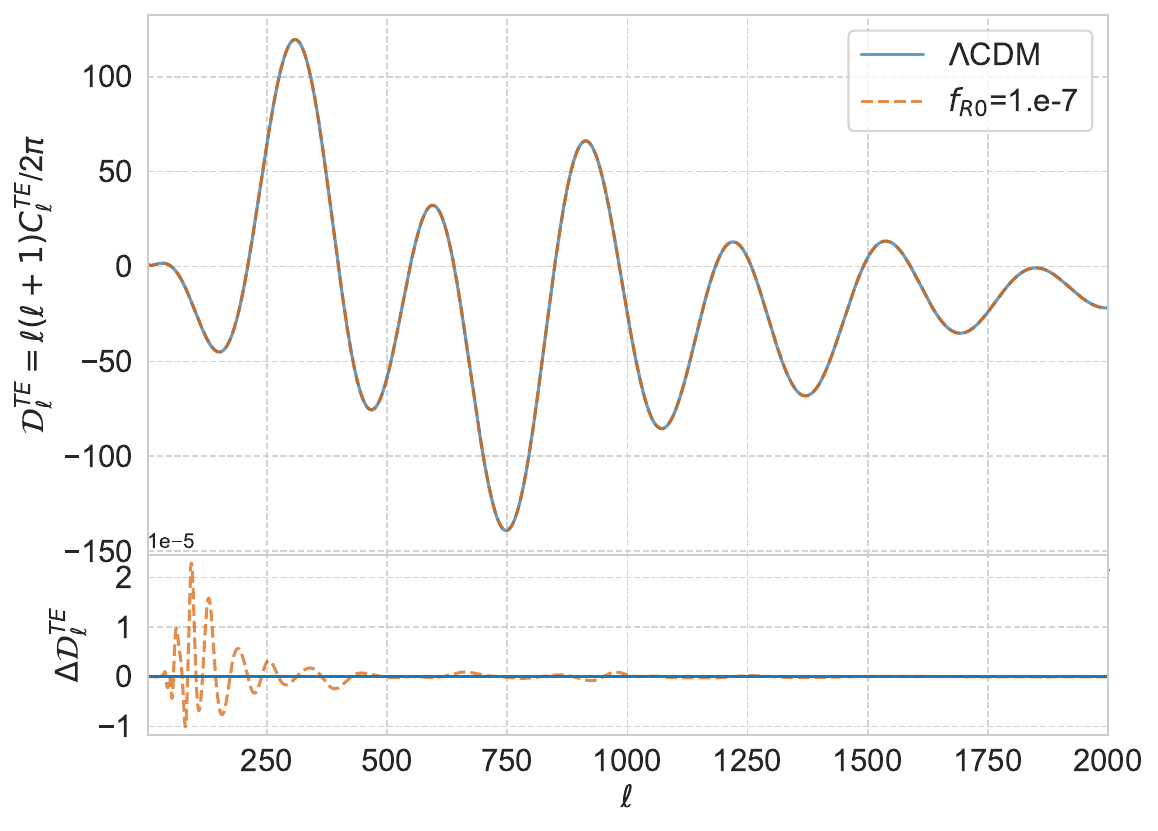}}
\subfloat{\includegraphics[width=0.525\linewidth]{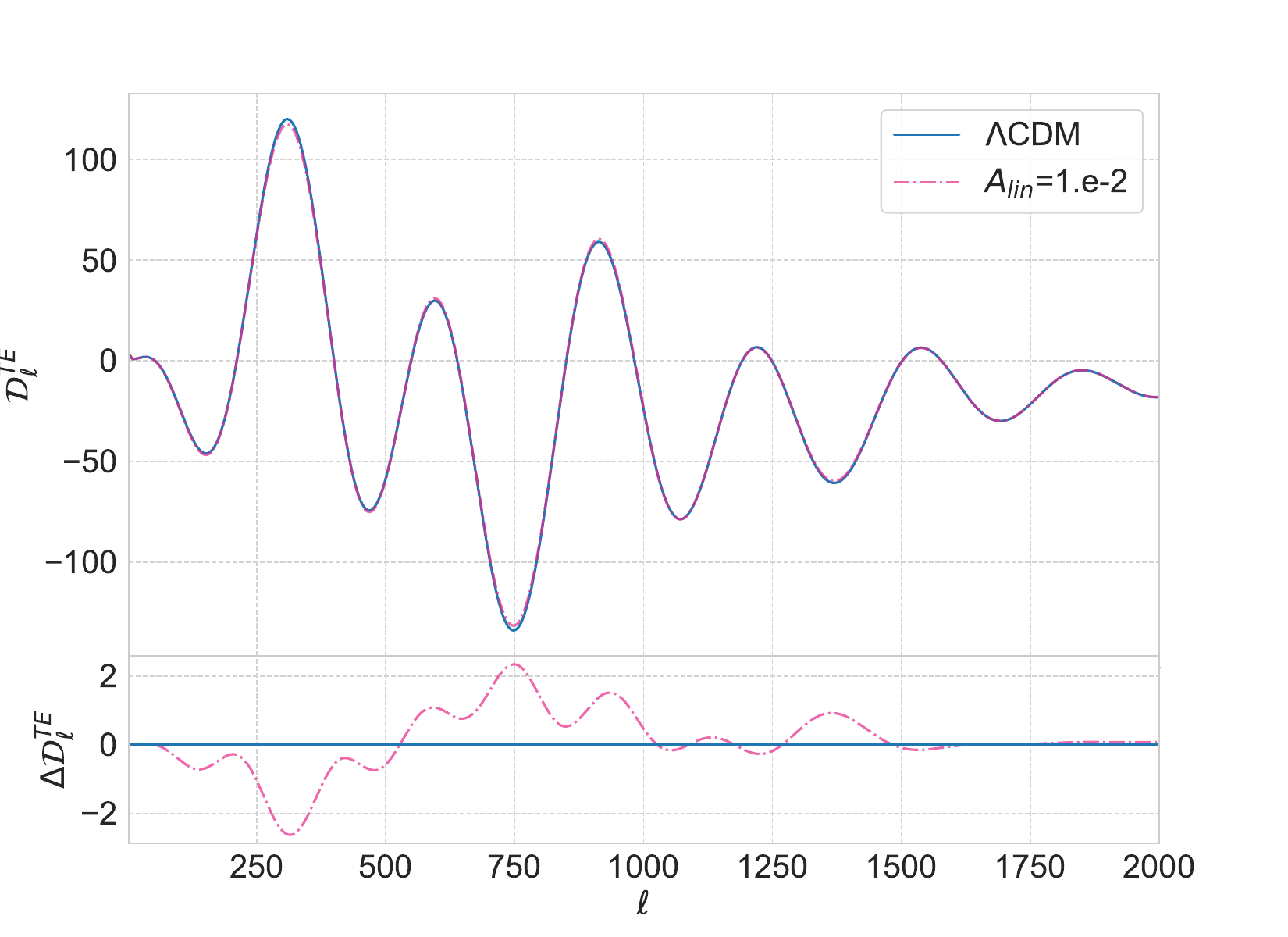}}\\
\subfloat{\includegraphics[width=0.5\linewidth]{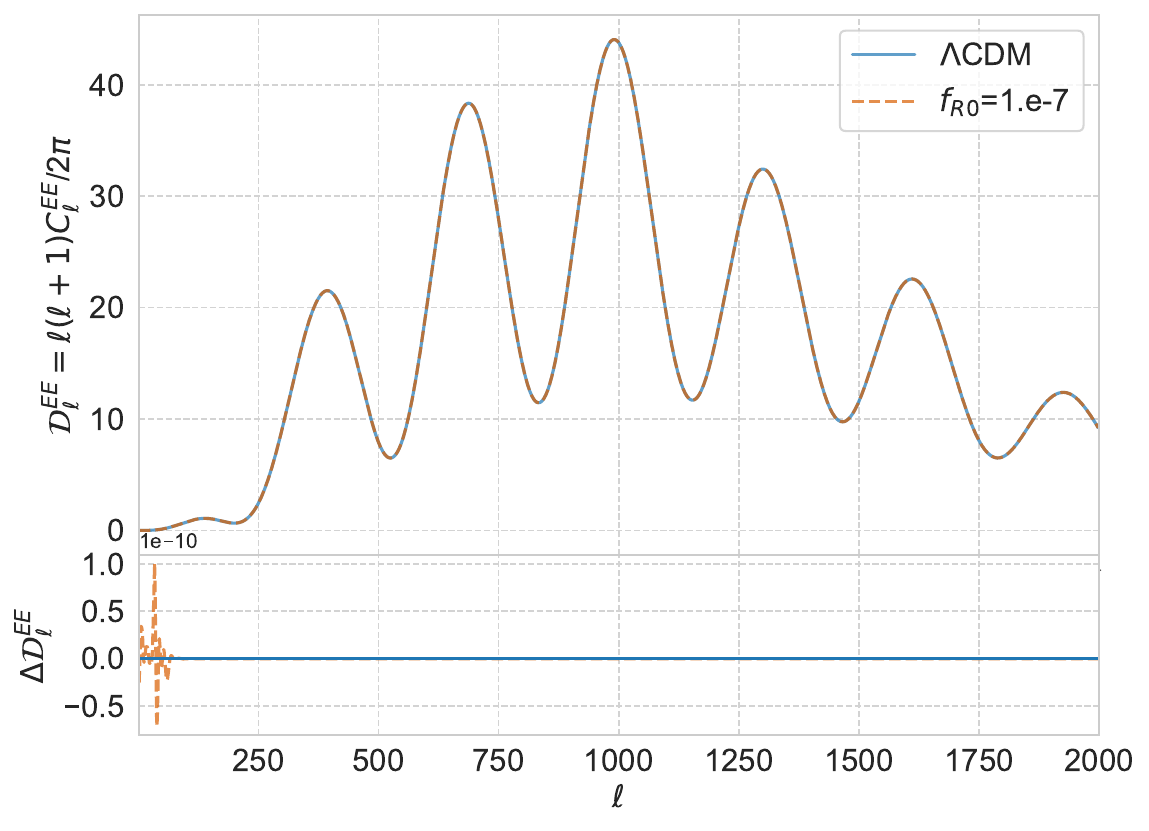}}
\subfloat{\includegraphics[width=0.525\linewidth]{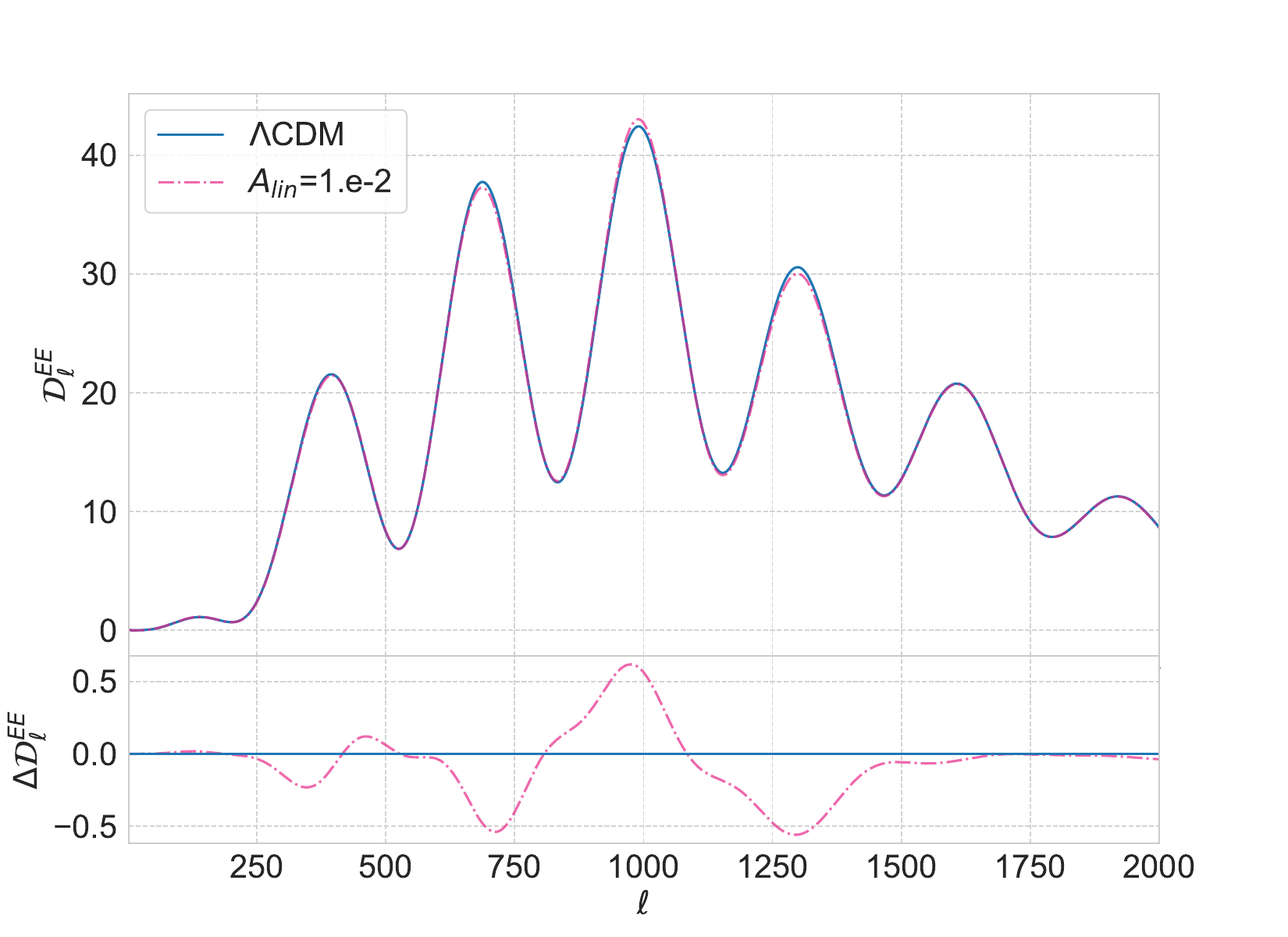}}
\caption{Left: $f_{R0}$ effect on the TT, TE and EE components. Right: Feature template on the primordial power spectrum, this is reflected on the effect of the $A_\mathrm{lin}$ parameter size on the TT, TE and EE components.}
\label{fig:FR0_Feat_Cls}
\end{figure}

\begin{figure}[H]
\centering
\subfloat{\includegraphics[width=0.4\linewidth]{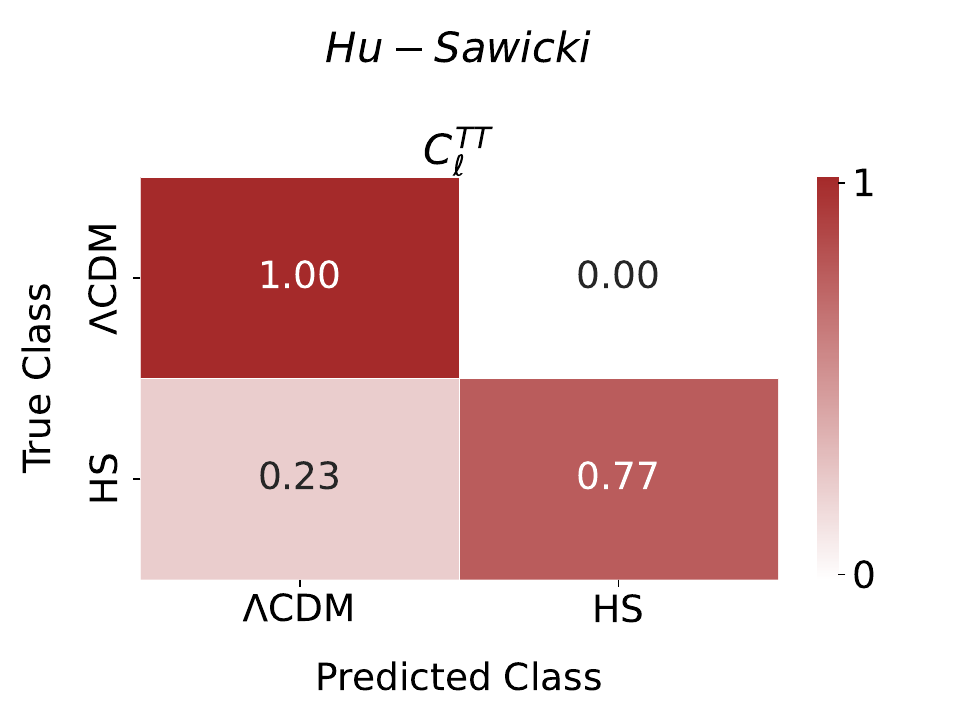}}
\subfloat{\includegraphics[width=0.4\linewidth]{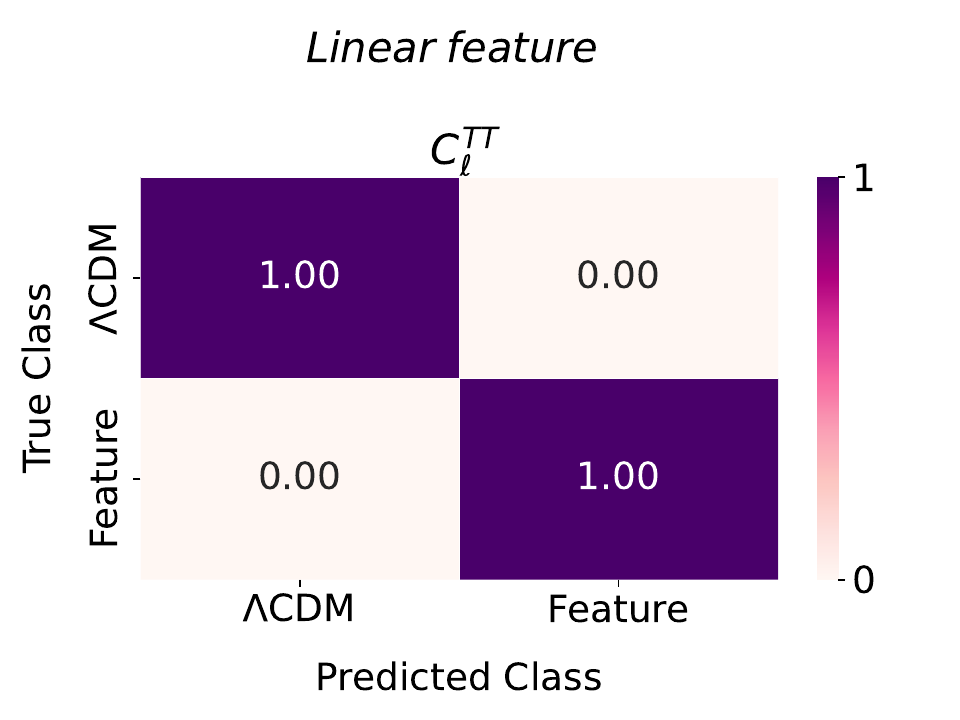}}\\
\subfloat{\includegraphics[width=0.4\linewidth]{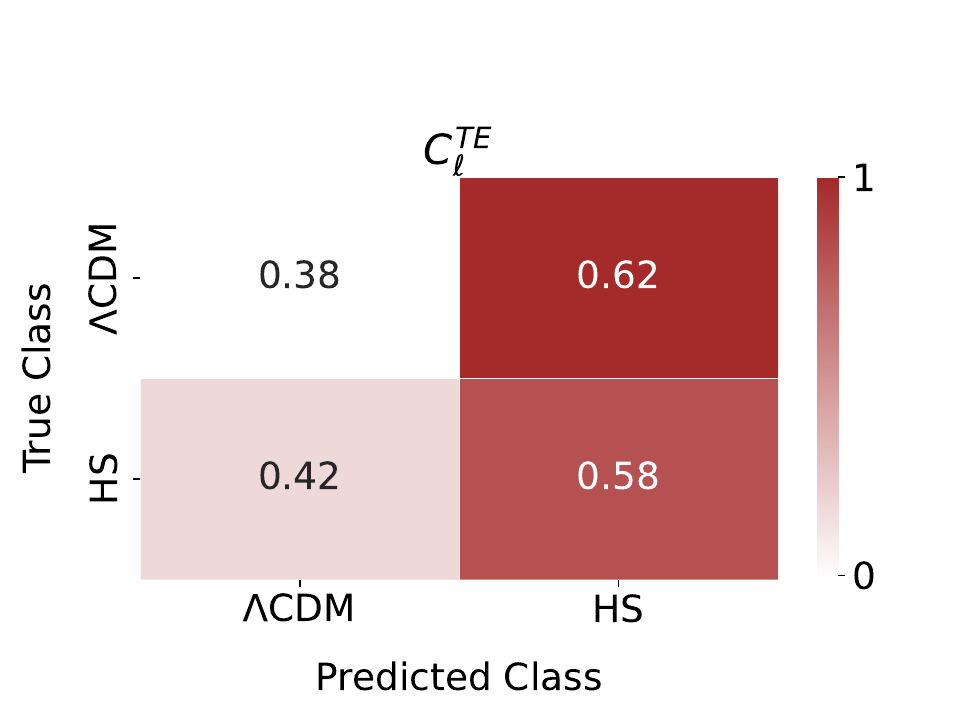}}
\subfloat{\includegraphics[width=0.4\linewidth]{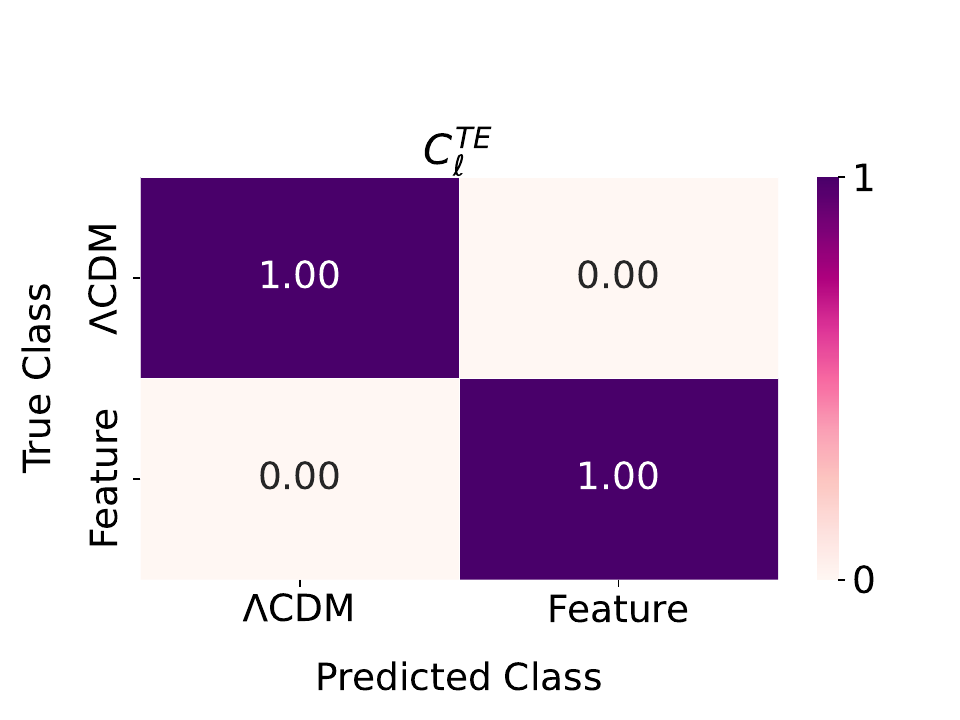}}\\
\subfloat{\includegraphics[width=0.4\linewidth]{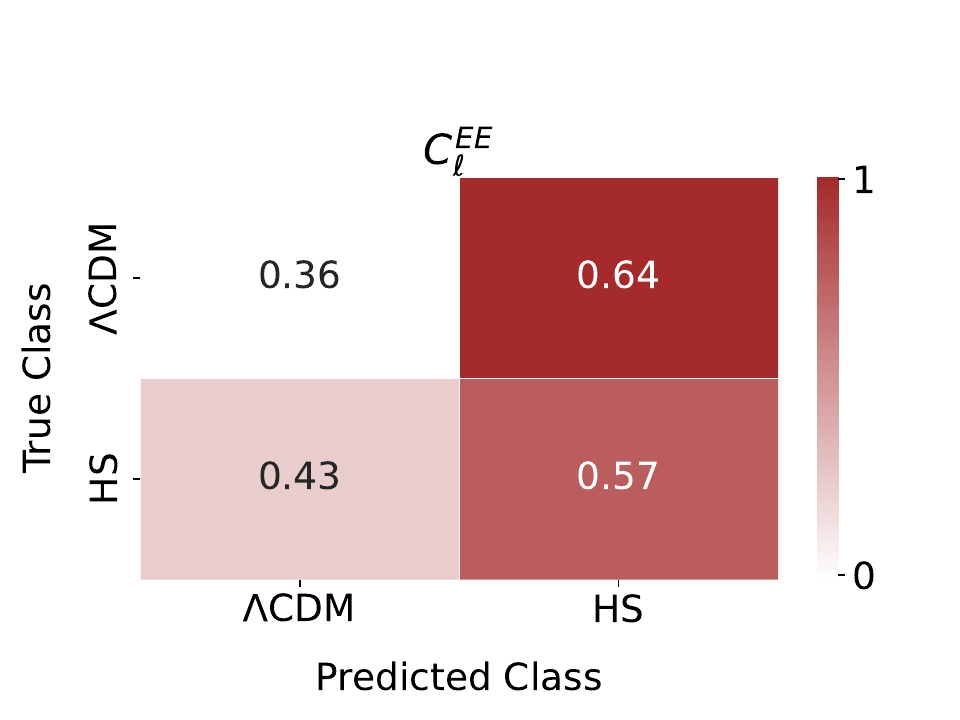}}
\subfloat{\includegraphics[width=0.4\linewidth]{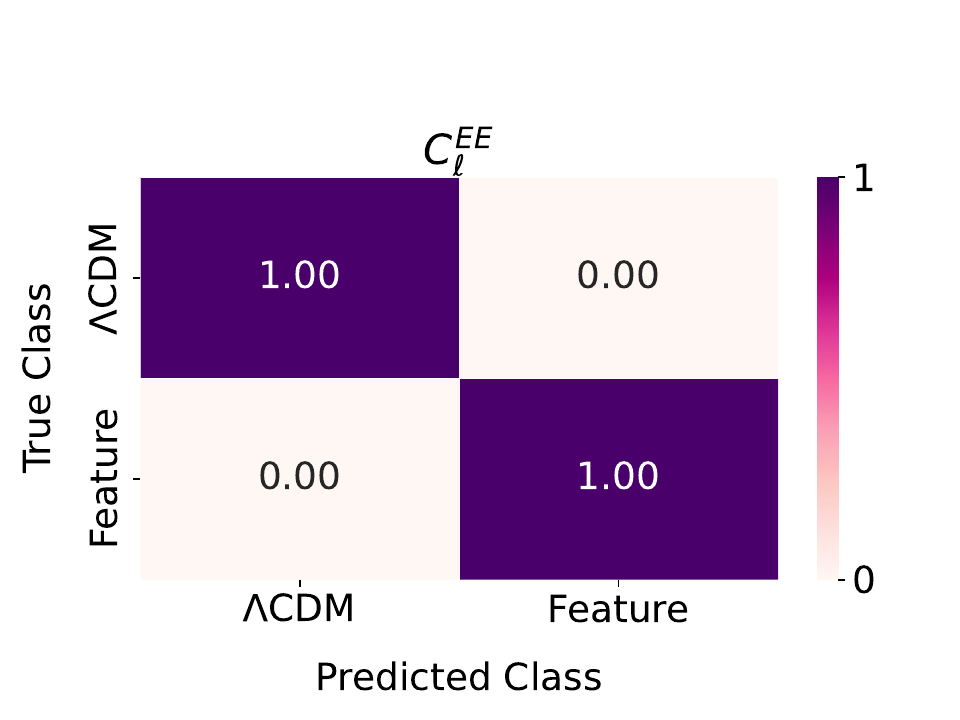}}\\
\subfloat{\includegraphics[width=0.4\linewidth]{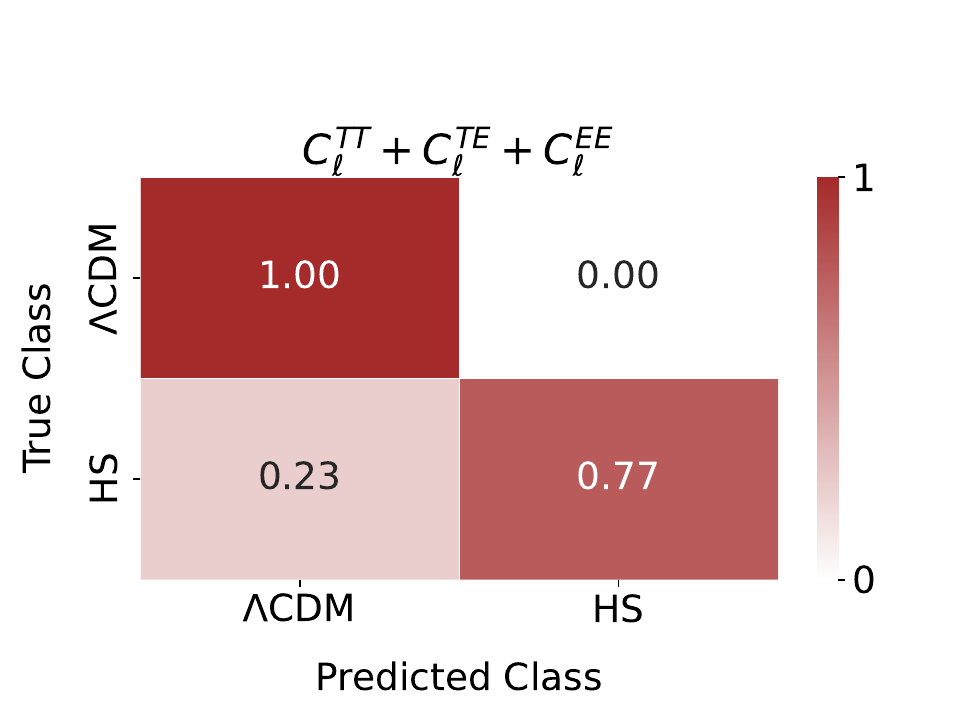}}
\subfloat{\includegraphics[width=0.4\linewidth]{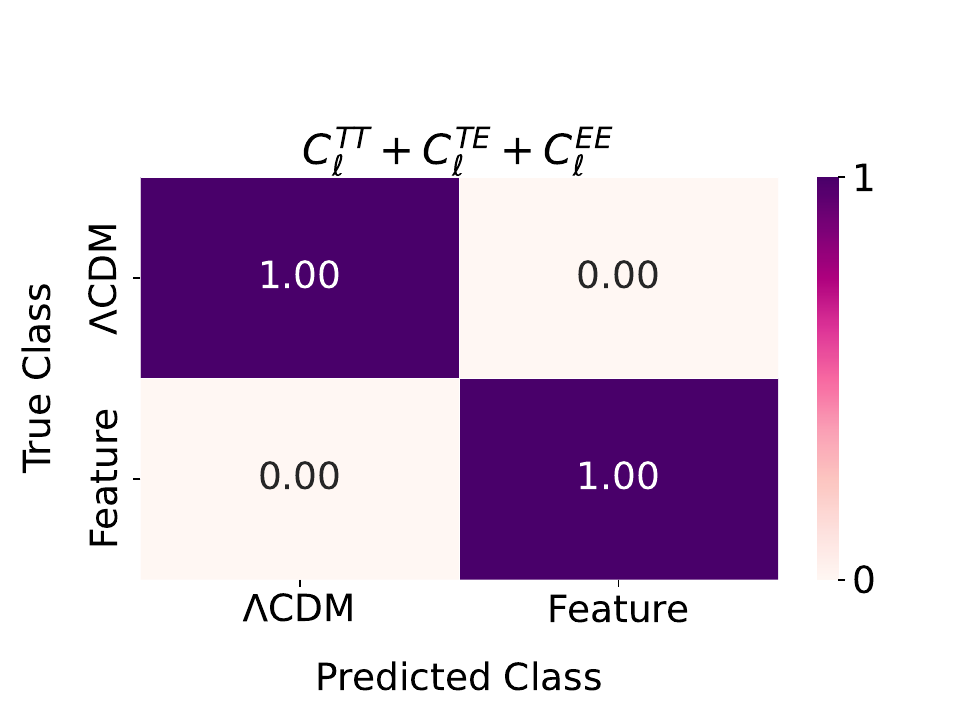}}\\
\caption{Confusion matrices of the neural network's performance. Results of the classification when trained and tested on each component of the temperature and polarization angular Power Spectrum, and its combined analysis. Two models were studied separately (against $\Lambda$CDM): for the early Universe case, a linearly-spaced template to search for primordial features in the primordial power spectrum (with a range of values of $A_{\operatorname{lin}} \in [10^{-2},5 \times 10^{-2}]$), and for the late-time Universe, the Hu-Sawicki model to test for deviations from $\Lambda$CDM (with a range of values of $|f_{R0}|\in[10^{-6},5\times 10^{-6}]$).}
\label{fig:confusion_matrices}
\end{figure}

In the context of MG, the NN performance in detecting $C_{\ell}^{TT}$ coming from $\Lambda$CDM or HS is 100$\%$ and 77$\%$ respectively (see \autoref{fig:confusion_matrices}), compared to the cases of the other components $C_{\ell}^{TE}$ and $C_{\ell}^{EE}$, where the overall performance drops to 47$\%$ and 48$\%$. When studying the case in which we combined all the data, we concluded that the NN is mostly learning to differentiate between models mainly from the temperature angular power spectrum component. To understand this, we illustrate in \autoref{fig:FR0_Feat_Cls} the effect of $f_{R0}$ on the $C_\ell$ components of the CMB, where only the $C_{\ell}^{TT}$ one is affected by the changes in this parameter. This is why the performance of the NN to classify between $\Lambda$CDM and HS is clearly better in the temperature angular power spectrum than the other two cases: $C_{\ell}^{TE}$ and $C_{\ell}^{EE}$ do not show any deviations in HS with respect to $\Lambda$CDM. 

We would like to highlight that we used 8 different architectures, 4 for each case, instead of training the same model over the data coming from different cases.

\begin{figure}[h!]
\centering
\subfloat{\includegraphics[width=0.52\linewidth]{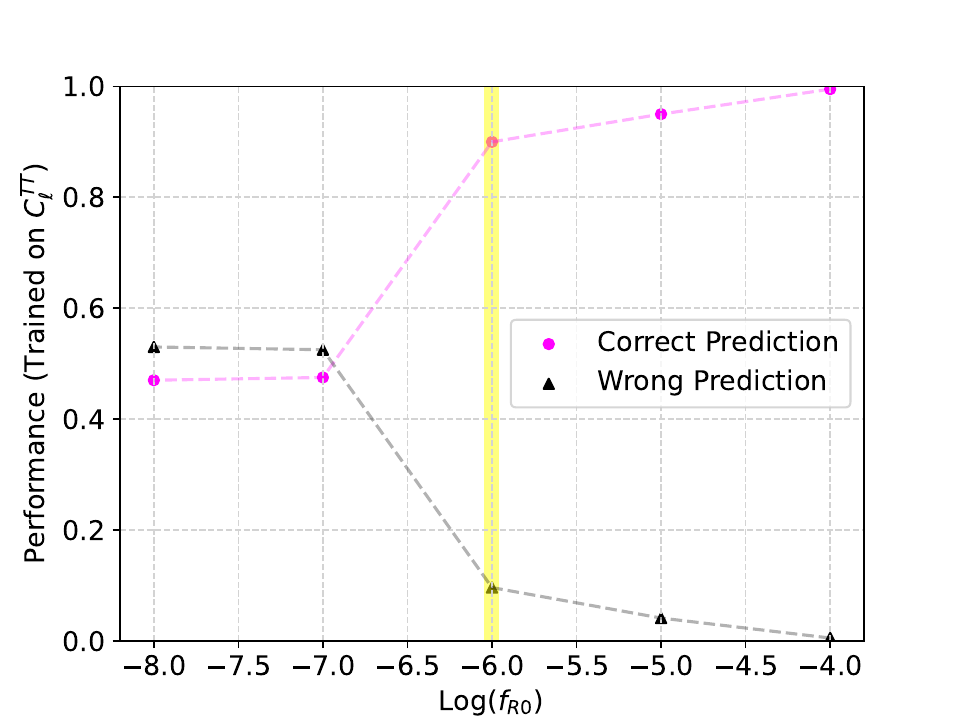}}
\subfloat{\includegraphics[width=0.52\linewidth]{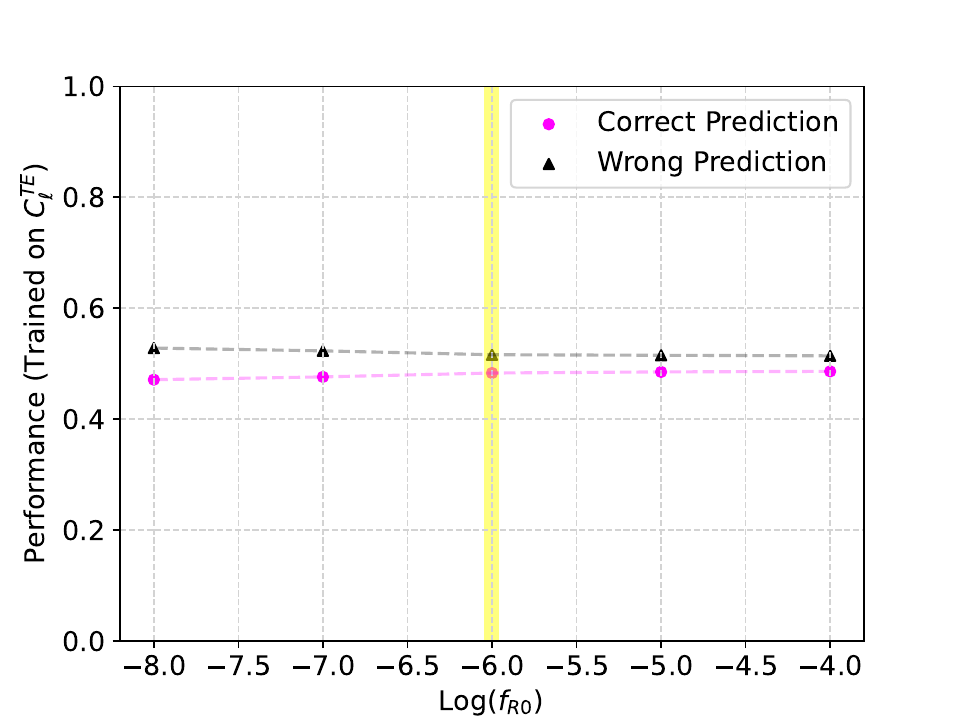}}\\
\subfloat{\includegraphics[width=0.52\linewidth]{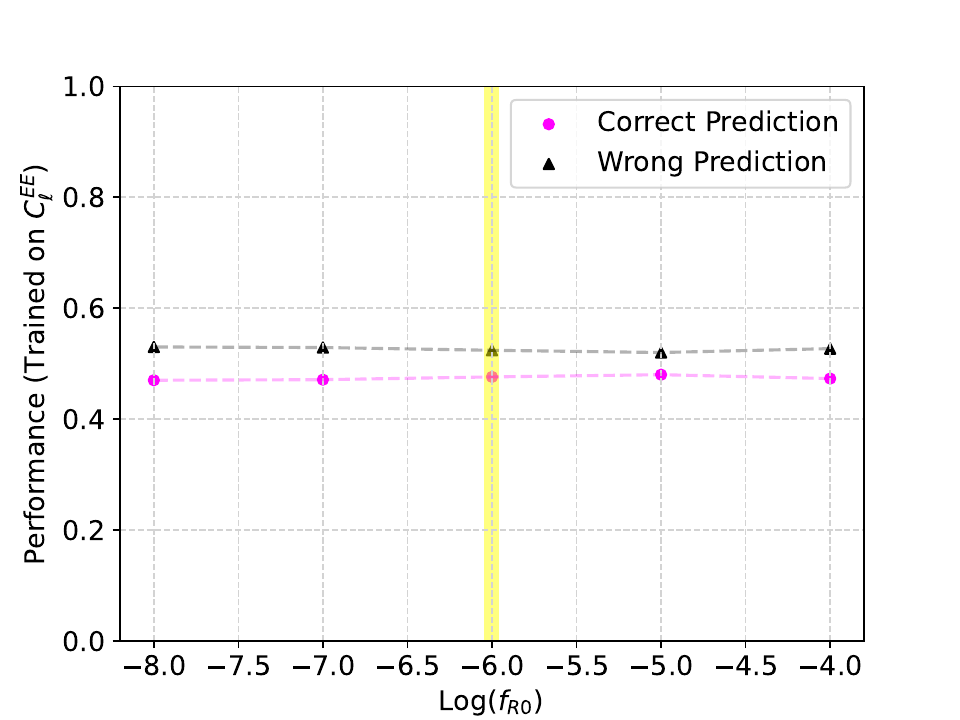}}
\subfloat{\includegraphics[width=0.52\linewidth]{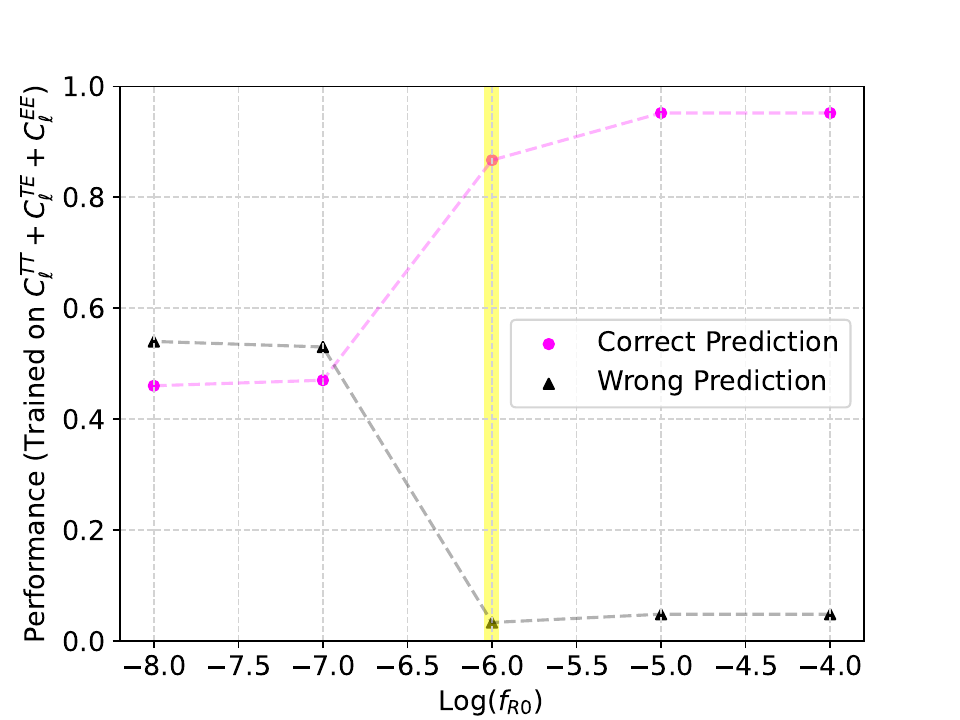}}\\
\caption{Semi-log plot of $f_{R0}$ values vs architecture performance for the components (up) $C_{\ell}^{TT}$, $C_{\ell}^{TE}$ (bottom) $C_{\ell}^{EE}$, joint analysis: $C_{\ell}^{TT}+C_{\ell}^{TE}+C_{\ell}^{EE}$. The yellow region corresponds to the fiducial value of $|f_{R0}|=10^{-6}$.}
\label{fig:FR0_performance}
\end{figure}

We also aimed to study the point (if any) in which the NNs performance breaks down or changes, so we generated datasets for different ranges of $f_{R0}$, and then compared the performance on each of them of them, and also the combined case $C_{\ell}^{TT}+C_{\ell}^{TE}+C_{\ell}^{EE}$. This information is available in \autoref{fig:FR0_performance}, where we can notice that for the $C_{\ell}^{TT}$ case, the performance breaks down for values of $|f_{R0}|=10^{-7}$, and for the other two modes, the NN cannot differentiate between models regardless of the value of this parameter. This value has proven to be still in agreement with current surveys (see, for example \cite{casas2023euclid}), while the $\Lambda$CDM model is recovered with $|f_{R0}|=0$. In fact, the NN was able to \textit{learn} a feature of the HS model: the HS model only shows deviations on the $C_{\ell}^{TT}$ component (see \cite{bellini2018comparison} where they verify that this is indeed the case by comparing Boltzmann solvers). This blind test strengthens the robustness of our pipeline, as the NN could not classify the data realization as neither $\Lambda$CDM nor HS using $C_{\ell}^{TE}$ and/or $C_{\ell}^{EE}$.

\subsection{Search for primordial features: linearly-spaced template}\label{subsec:Featuremodel}
With respect to the case for features in the primordial power spectrum, defined as linearly-spaced, the NN is able to differentiate between models in all three angular power spectra, $C_{\ell}^{TT}$, $C_{\ell}^{TE}$ and, $C_{\ell}^{EE}$, and also for the combined case. In \autoref{fig:confusion_matrices}, we notice that the performance is 1 for all the cases, which is sensible if we see the effect of $A_{\text{lin}}$ on the modes in \autoref{fig:FR0_Feat_Cls} for the fiducial value of $A_{\text{lin}}=10^{-2}$, where clear deviations are shown with respect to the $\Lambda$CDM model. This is expected since primordial features in the power spectrum are expected to introduce changes on all CMB $C_\ell$ components, as given by \autoref{eq:C_ells}. 

\begin{figure}[h!]
\centering
\subfloat{\includegraphics[width=0.52\linewidth]{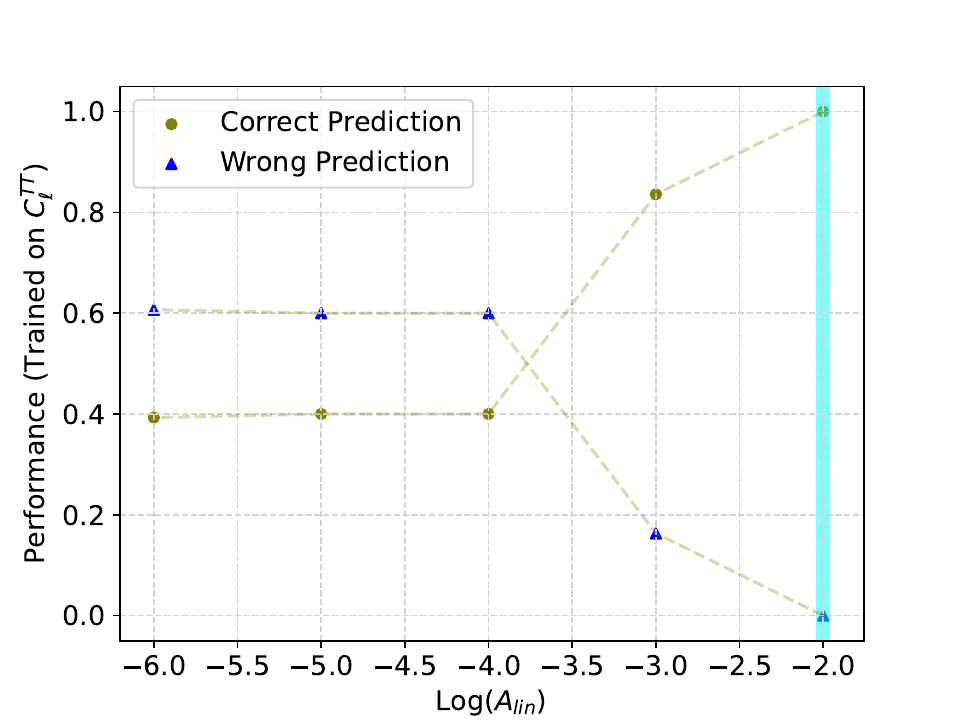}}
\subfloat{\includegraphics[width=0.52\linewidth]{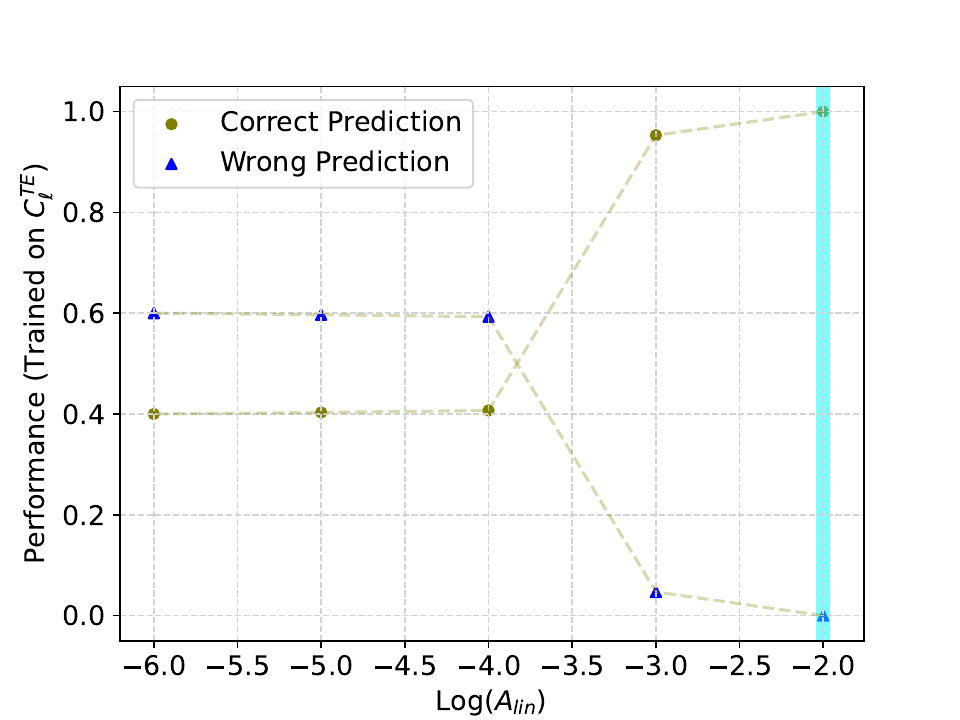}}\\
\subfloat{\includegraphics[width=0.52\linewidth]{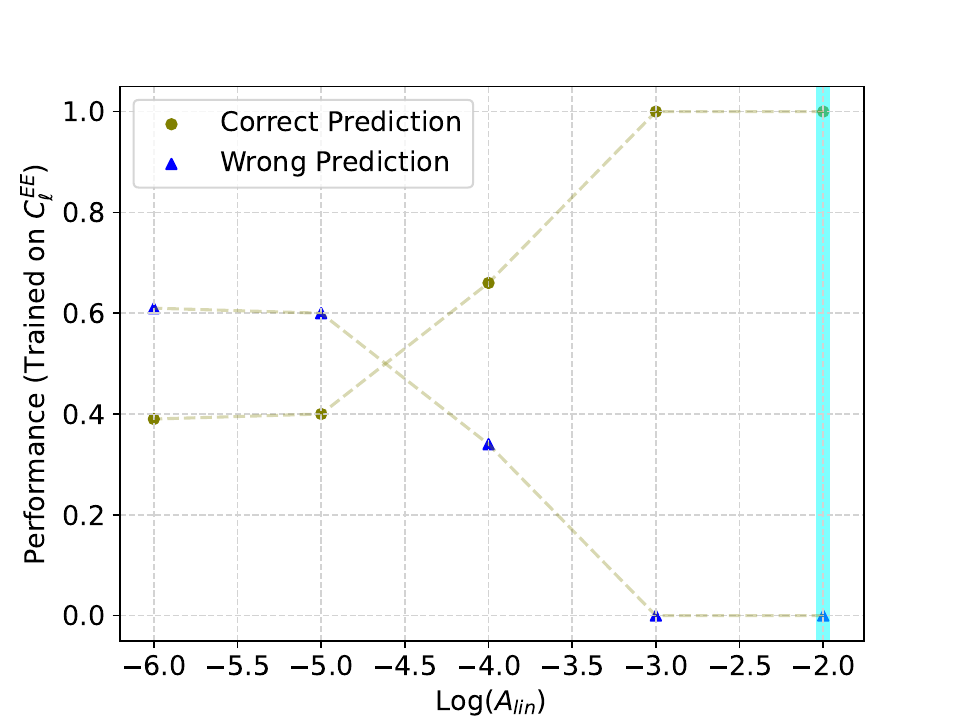}}
\subfloat{\includegraphics[width=0.52\linewidth]{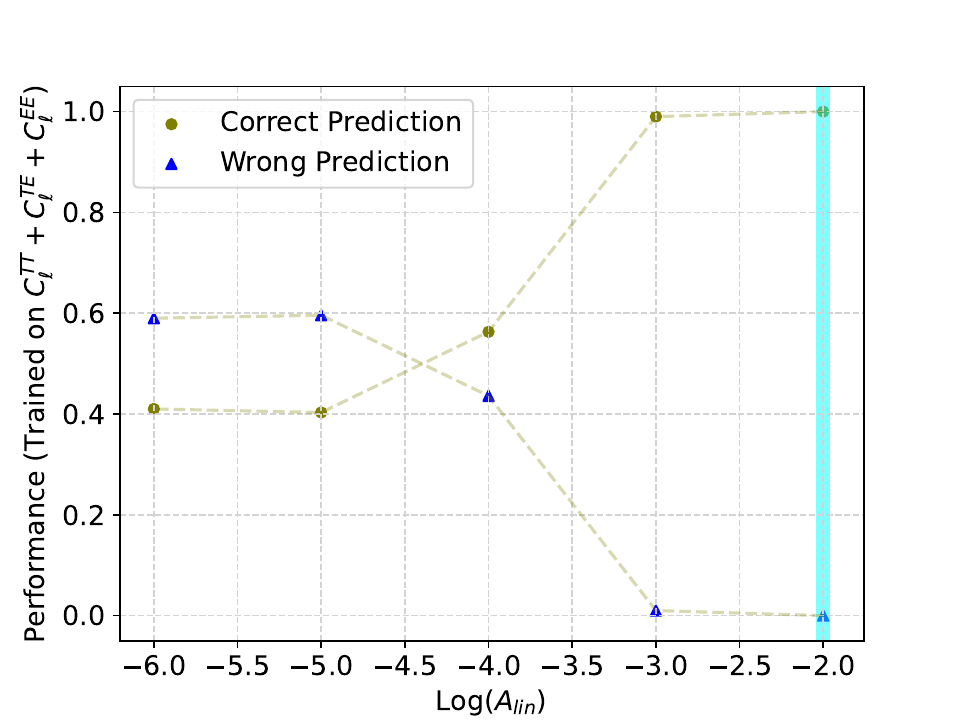}}
\caption{Semi-log plot of $A_{\text{lin}}$ values vs architecture performance for the components (up) $C_{\ell}^{TT}$, $C_{\ell}^{TE}$ (bottom) $C_{\ell}^{EE}$, joint analysis: $C_{\ell}^{TT}+C_{\ell}^{TE}+C_{\ell}^{EE}$. The cyan region corresponds to the fiducial value of $A_{\text{lin}}=10^{-2}$.}
\label{fig:Ax0_performance}
\end{figure}

For completion, we also analyzed the point in which the NN's performance breaks down, and for all the cases there is similar value of $A_{\text{lin}}=10^{-4}$, where the correct prediction decreases to 0.4 for $C_{\ell}^{TT}$ and $C_{\ell}^{TE}$ and to $\approx0.6$ for $C_{\ell}^{EE}$ and the combined case (see \autoref{fig:Ax0_performance}). In general, the NN is able to discern between $\Lambda$CDM and the linearly-spaced feature model for fiducial values of $A_{\text{lin}}$. This is a highlight of our pipeline, since the NN is able to classify well even when the differences between the models are very small. 

\subsection{ML interpretability and assessment tests}\label{subsec:Interpretability}

\begin{figure}[h!]
\raggedright
\subfloat{\includegraphics[width=0.97\linewidth]{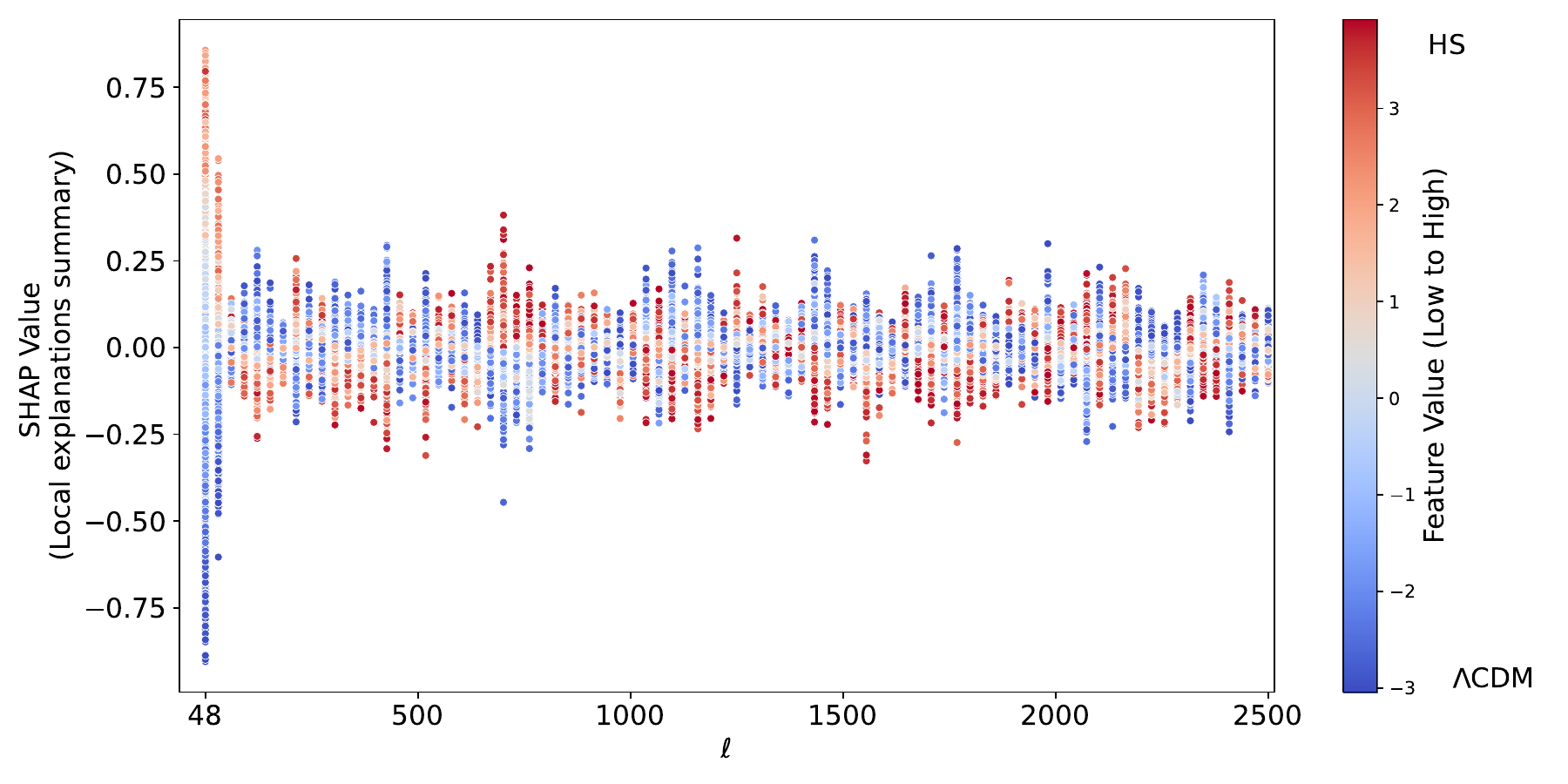}}\\
\subfloat{\includegraphics[width=0.82\linewidth]{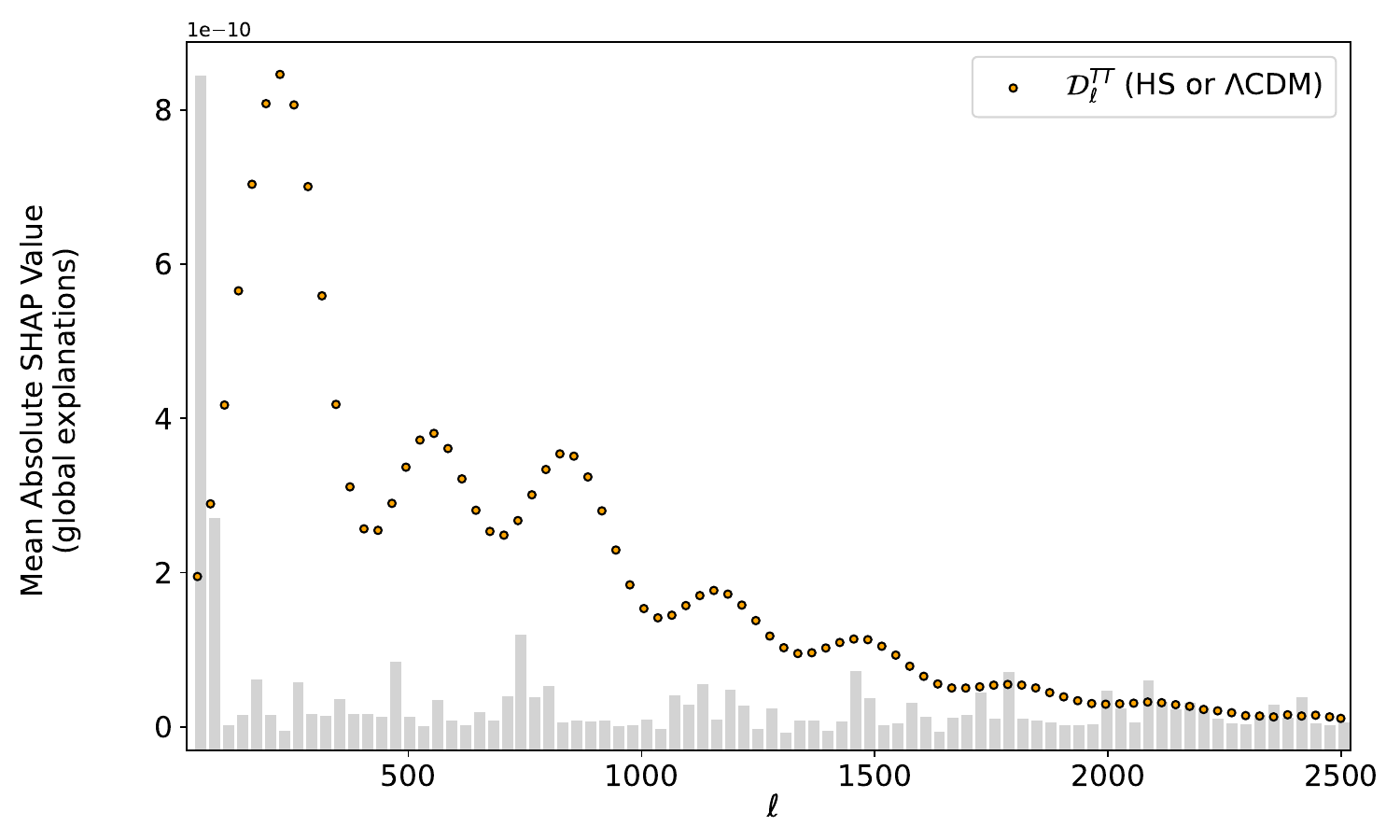}}
\caption{Machine Learning explanations for the Hu-Sawicki model. Top: local interpretability given by the \texttt{SHAP} values (high value, Hu-Sawicki and low value, $\Lambda$CDM). Bottom: global interpretability by the mean absolute \texttt{SHAP} values that corresponds to each $C_{\ell}^{TT}$.}
\label{fig:interpretability_HS}
\end{figure}
Machine Learning (ML) interpretability has become essential in deep learning, as understanding how these ML techniques reach their conclusions is necessary for building trust in their application across various fields, particularly in analyzing complex datasets \cite{zhang2021survey}. To assess how the NN learn, we focused our work on the interpretability tool known as \texttt{SHAP} (SHApley Additive exPlanations)\footnote{\url{https://github.com/SHAP/SHAP.git}}. In general, ML interpretability follows two approaches: local and global explanations. Global explanations provide an overview of the learning process throughout the entire dataset, whereas the local method focuses on studying a region in the hyperparameter space. \texttt{SHAP}, which can be employed for both local and global explanations, is a method for explaining the output of a NN architecture by attributing the contribution of each element to the final prediction, which is known as \textit{feature importance}\footnote{Not to be confused with the feature template of the primordial power spectrum in extensions beyond single-field inflation; therefore in the body of the text of this work, we will refer to it as "element importance".}.

\texttt{SHAP} is based on cooperative game theory, and helps to understand the most important input elements (or "players") by computing an adapted version of the so-called \textit{Shapley values} \cite{lundberg2020local}. It has a bottom-up approach: for the local explanations, it assesses the contribution of each input datapoint (in our case, the ${C}_{\ell}$) to the prediction, by considering all possible combinations. In this context, Shapley values are calculated by introducing each element $x$ (one by one) into a conditional function of the model's output
\begin{equation}
    f_x(S)\approx E\left[f(x) \mid x_S\right],
\end{equation}
which can be understood as the expected prediction given that only the elements in a subset $S$ are known. The change in the model's decision at each step is attributed to the newly introduced element. This process is then averaged across all possible element orderings to ensure a fair distribution of contributions. Hence, the \texttt{SHAP} values produces attributions $\phi_i(f, x)$ that matches the original model output $f(x)$ \cite{lundberg2020local}, given by,  
\begin{equation}
\phi_i(f, x)=\sum_{m \in \mathcal{M}} \frac{1}{N!}\left[f_x\left(P_i^m \cup i\right)-f_x\left(P_i^m\right)\right],
\end{equation}
\noindent where $\mathcal{M}$ is the set of all feature orderings, $P_i^m$ is the set of all elements that come before the $i^{th}$ element in ordering $m$, and $N$ is the number of input elements for the model. For a revision on this topic refer to \cite{lundberg2020local, sundararajan2020many} and \cite{ghorbani2019data}. On the contrary, for the global interpretability approach, this method calculates the mean absolute \texttt{SHAP} values, giving overall importance to each element for the model's outcome.

In this work, \texttt{SHAP} is used for one of the first times as interpretability tool in cosmological model selection and interpretability \cite{Valle:2020jyk, SHAPphotometry}. We employed \texttt{SHAP} for both, local and global interpretability assessments for the temperature angular power spectrum $C_\ell^{TT}$ for the two extension models. The analysis takes approximately 3 hours for each model. For the case of the modified gravity model, we can see the Hu-Sawicki NN explanations in \autoref{fig:interpretability_HS}; the upper panel displays the local interpretability assessment with \texttt{SHAP} values that correspond to each $\mathcal{D}_{\ell}^{TT}$. Higher values mean that their corresponding elements are more likely to contribute to a Hu-Sawicki outcome, and lower values to a $\Lambda$CDM one. In the lower panel, we see the global interpretability results with the mean absolute \texttt{SHAP} values, that correspond to each $\mathcal{D}_{\ell}^{TT}$. The dots represent one randomly selected realization of the data (either belonging to the HS model or to $\Lambda$CDM). We observe how the first $\mathcal{D}^{TT}_\ell$ values are the critical one for the NN to learn the main elements to classify either $\Lambda$CDM or HS. This is expected as observed in \autoref{fig:FR0_Feat_Cls}, where the main changes with respect to $\Lambda$CDM for the HS model take place at low multipole values.

\begin{figure}[h!]
\raggedright
\subfloat{\includegraphics[width=0.97\linewidth]{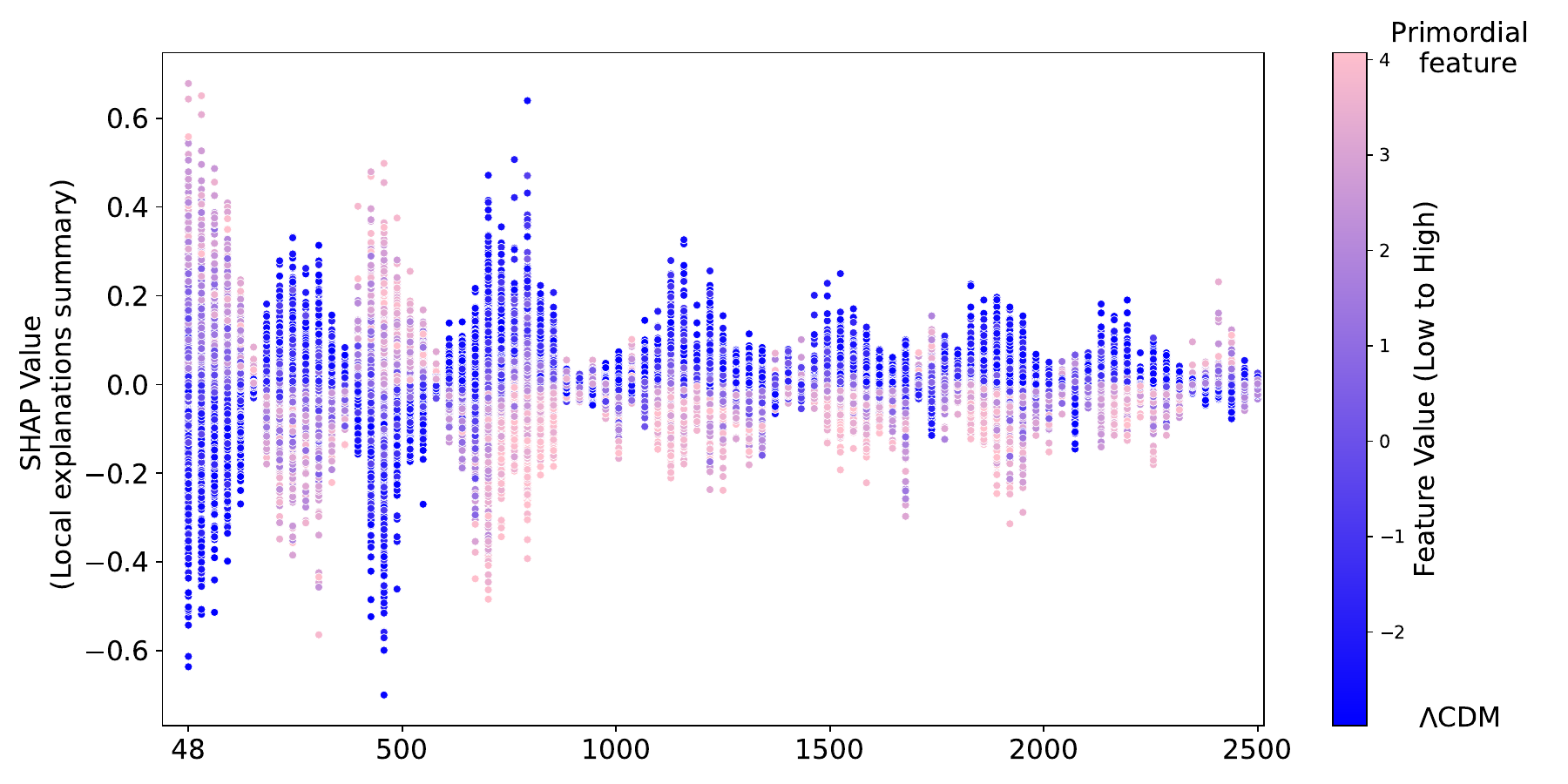}}\\
\subfloat{\hspace{6pt}\includegraphics[width=0.80\linewidth]{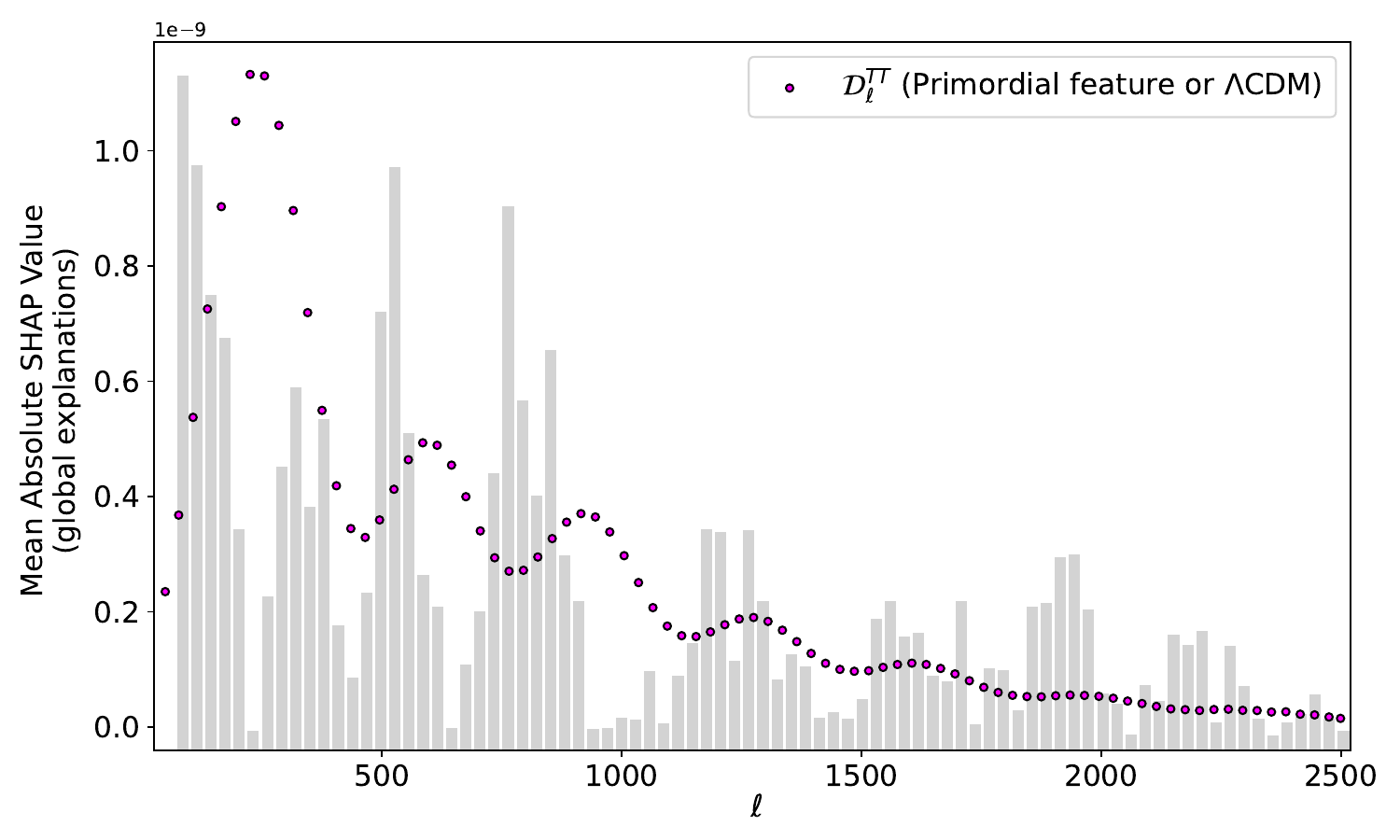}}
\caption{Machine Learning explanations for the linearly-spaced primordial feature model. Top: local interpretability given by the \texttt{SHAP} values (high values, primordial feature model and low ones, $\Lambda$CDM). Bottom: global interpretability by the mean absolute \texttt{SHAP} values that corresponds to each $C_{\ell}^{TT}$.}
\label{fig:interpretability_feature}
\end{figure}

Subsequently, in figure \autoref{fig:interpretability_feature}, we show the explanations for the linearly-spaced primordial feature template model; as in the previous case, the upper panel displays the local interpretability part with \texttt{SHAP} values that correspond to each $\mathcal{D}_{\ell}^{TT}$. In this case, higher values show that their corresponding elements are more likely to contribute to a feature-template model outcome, and lower values to a $\Lambda$CDM one. In the same way, the lower panel displays the global interpretability conclusions, with the mean absolute \texttt{SHAP} values, that correspond to each $\mathcal{D}_{\ell}^{TT}$. Again, the dots represent one randomly selected realization of the data (either belonging to the feature-template model or to $\Lambda$CDM). In this case, both the local and global explanations trace properly the changes introduced in the $\mathcal{D}_{\ell}^{TT}$ due to the primordial feature template in \autoref{fig:Feature}, identifying as critical those $\ell$ where the corresponding Planck 18 uncertainty is larger, and therefore, more likely to allow for deviations from the power-law primordial power spectrum. 
In summary, we notice that the NN are properly learning how to differentiate models based on the deviations displayed in the lower panels of \autoref{fig:FR0_Feat_Cls}, because it seems that the peaks of the most important elements in both \autoref{fig:interpretability_HS} and \autoref{fig:interpretability_feature} coincide with these deviations. We also raise an interesting point by revisiting \autoref{fig:confusion_matrices}, where we conclude that NN perform better in differentiating the primordial feature template from $\Lambda$CDM than the case of Hu-Sawicki. In light of the \texttt{SHAP} values, this is expected because the primordial feature model displays more numerous important elements than those of the HS analysis, helping the NN identify the key-elements for classification.

\section{Conclusions}\label{sec:conclusions}

In this work, we presented our Machine Learning pipeline designed for model selection using deep Neural Networks to classify two models beyond the Standard Cosmological Model. Our first focus was on late-time physics, specifically deviations from General Relativity within the framework of the Hu-Sawicki $f(R)$ model. The second focus was on early universe hypotheses, particularly the search for linearly-spaced features in the primordial power spectrum. The architecture of our deep Neural Network pipeline has been optimized to maximize the learning of key features while minimizing over-fitting.

To train our pipeline, we generated Planck 2018-like data, including its specifications and uncertainties for the mentioned cases. We tested its ability to distinguish between data derived from the Hu-Sawicki model vs $\Lambda$CDM and the primordial feature model vs $\Lambda$CDM. Our pipeline comprises a total of eight Neural Networks: four dedicated to the three polarization modes and one that combines all of them for testing the linearly spaced primordial feature model, as well as four for evaluating the Hu-Sawicki model. Generally, the architectures remained consistent, with the primary variation being the number of neurons in the input layer, which corresponded to the number of binned Planck 2018 angular power spectrum data points.

The results demonstrate that our pipeline effectively learned the crucial elements necessary to classify the test dataset. Specifically, for the Hu-Sawicki datasets (for a value of $\left|f_{R 0}\right|=10^{-6}$), the Neural Network achieved a correct prediction rate of 93\% for the temperature power spectrum. However, for polarization and cross-data, the performance dropped to 0.47 and 0.48, respectively, aligning with theoretical expectations, as the Hu-Sawicki model does not exhibit differences compared to $\Lambda$CDM in the polarization CMB angular power spectra. This outcome indicates that our Neural Network accurately identified this insensitivity and successfully passed this sanity check. In the second scenario, we aimed to differentiate between the primordial linearly-spaced feature model and $\Lambda$CDM across temperature, polarization, and cross-data (as well as a combination of all three) for an oscillation amplitude value of ${A}_{\text {lin }}=0.01$. Remarkably, the pipeline achieved a 100\% correct prediction rate across all components, making this kind of ML pipelines highly useful in the community to assess the potential of a dataset to constrain features in the primordial power spectrum. In fact, this computationally cheaper approach can be applied before addressing a full Bayesian inference exercise to sample the probability posterior distributions of the features' parameters to actually discern the capacity to detect a possible primordial feature template.

To evaluate and understand how our pipeline is learning to classify amongst models, we implemented \texttt{SHAP} as an interpretability tool to extract meaningful insights about the main input elements that drive these Neural Networks to certain outcomes. The application of this particular interpretability tool is one of the first applications of \texttt{SHAP} to cosmological model selection. We confirmed that our architectures can differentiate between models by focusing on the elements of the angular power spectra that reflect the deviations between the proposed alternatives and $\Lambda$CDM, these key elements were identified prior to training. We conclude with highlighting our ML interpretability results, since feature importance is crucial for evaluating the learning process and deep learning algorithms do not always make decisions based on the features or elements we intend for them to learn. 

A point worth highlighting of our work is that we only trained the NN on grids of one or two parameters, namely $\Omega_{\text{cdm}}$ for the $\Lambda$CDM model, $\Omega_{\text{cdm}}+f_{R0}$ for the Hu-Sawicki model and $\Omega_{\text{cdm}}+A_{\rm lin}$ for the primordial features one. However, to perform a fair comparison between the real CMB Planck data against the underlying cosmological model we require at least 6 parameters: $\Omega_{\text{cdm}}$, $\Omega_{b}$, $n_s$, $A_s$, $\tau$ and $H_0$ (the minimal $\Lambda$CDM model), and possibly also the curvature parameter $\Omega_k$ to allow for different geometrical configurations. For this reason, we did not test our architecture with measured Planck 18 angular power spectra and will address this question in future work, where we will expand the scope of this analysis to a more in-depth comparison with the real data by training the NN on a multi-dimensional grid. The main reason for this decision was to maintain the focus on the interpretability of the network and show a robust proof-of-concept for the application of \texttt{SHAP}.

Additionally to the aforementioned and looking ahead to future work, we see the potential for extending our study by applying the pipeline to classify models using CMB temperature and polarization maps, leveraging techniques like Convolutional Neural Networks or Graph Neural Networks. Additionally, we aim to apply the pipeline to other summary statistics from various cosmological observables, such as cosmic shear in Large-Scale Structure and photometric angular clustering power spectra.

\acknowledgments
We thank Dr. Stéphane Ilic for interesting discussions about the Planck 18 covariance matrices. IO thanks ESTEC/ESA for the warm hospitality during the execution of this project, and for support from the \href{https://www.cosmos.esa.int/web/esdc/visitor-programme}{ESA Archival Research Visitor Programme}. IO thanks to Elena Donini for useful discussions during the ESA internship. IO and SN acknowledge support from the research project PID2021-123012NB-C43 and the Spanish Research Agency (Agencia Estatal de Investigaci\'on) through the Grant IFT Centro de Excelencia Severo Ochoa No CEX2020-001007-S, funded by MCIN/AEI/10.13039/501100011033. IO is also supported by the fellowship LCF/BQ/DI22/11940033 from ``la Caixa” Foundation (ID 100010434). GCH acknowledges support through the ESA research fellowship programme. SN and IO acknowledge the use of the Finis Terrae III Supercomputer which was financed by the Ministry of Science and Innovation, Xunta de Galicia and ERDF (European Regional Development Fund).

\bibliography{references.bib}

\providecommand{\noopsort}[1]{}\providecommand{\singleletter}[1]{#1}%
\begin{thebibliography}{10}

\bibitem{barreiro2010overview}
RB~Barreiro.
\newblock An overview of the current status of cmb observations.
\newblock {\em Highlights of Spanish Astrophysics V}, pages 93--102, 2010.

\bibitem{baumann2009tasi}
Daniel Baumann.
\newblock Tasi lectures on inflation.
\newblock {\em arXiv preprint arXiv:0907.5424}, 2009.

\bibitem{hu2008lecture}
Wayne Hu.
\newblock Lecture notes on cmb theory: From nucleosynthesis to recombination.
\newblock {\em arXiv preprint arXiv:0802.3688}, 2008.

\bibitem{collaboration2014planck}
Planck Collaboration, PAR Ade, N~Aghanim, C~Armitage-Caplan, M~Arnaud, M~Ashdown, F~Atrio-Barandela, J~Aumont, C~Baccigalupi, AJ~Banday, et~al.
\newblock Planck 2013 results. xxii. constraints on inflation.
\newblock {\em 0004-6361}, 571:22, 2014.

\bibitem{ade2016planck}
Peter~AR Ade, N~Aghanim, M~Arnaud, M~Ashdown, J~Aumont, C~Baccigalupi, AJ~Banday, RB~Barreiro, Nicola Bartolo, E~Battaner, et~al.
\newblock Planck 2015 results-xiv. dark energy and modified gravity.
\newblock {\em Astronomy \& Astrophysics}, 594:A14, 2016.

\bibitem{aghanim2020planck}
Nabila Aghanim, Yashar Akrami, Mark Ashdown, Jonathan Aumont, Carlo Baccigalupi, Mario Ballardini, Anthony~J Banday, RB~Barreiro, Nicola Bartolo, S~Basak, et~al.
\newblock Planck 2018 results-vi. cosmological parameters.
\newblock {\em Astronomy \& Astrophysics}, 641:A6, 2020.

\bibitem{akrami2020planck}
Yashar Akrami, Frederico Arroja, M~Ashdown, J~Aumont, Carlo Baccigalupi, M~Ballardini, Anthony~J Banday, RB~Barreiro, Nicola Bartolo, S~Basak, et~al.
\newblock Planck 2018 results-x. constraints on inflation.
\newblock {\em Astronomy \& Astrophysics}, 641:A10, 2020.

\bibitem{smoot1992structure}
George~F. Smoot, Charles~L. Bennett, Alan Kogut, Edward~L. Wright, Radek Stompor, et~al.
\newblock Structure in the cobe differential microwave radiometer first-year maps.
\newblock {\em The Astrophysical Journal Letters}, 396:L1--L5, 1992.

\bibitem{spergel2003first}
David~N Spergel, Licia Verde, Hiranya~V Peiris, Eiichiro Komatsu, MR~Nolta, Charles~L Bennett, Mark Halpern, Gary Hinshaw, Norman Jarosik, Alan Kogut, et~al.
\newblock First-year wilkinson microwave anisotropy probe (wmap)* observations: determination of cosmological parameters.
\newblock {\em The Astrophysical Journal Supplement Series}, 148(1):175, 2003.

\bibitem{hinshaw2013nine}
Gary Hinshaw, D~Larson, Eiichiro Komatsu, David~N Spergel, CLaa Bennett, Joanna Dunkley, MR~Nolta, M~Halpern, RS~Hill, N~Odegard, et~al.
\newblock Nine-year wilkinson microwave anisotropy probe (wmap) observations: cosmological parameter results.
\newblock {\em The Astrophysical Journal Supplement Series}, 208(2):19, 2013.

\bibitem{ade2011planck}
Peter~AR Ade, Nabila Aghanim, Monique Arnaud, Mark Ashdown, J~Aumont, C~Baccigalupi, A~Balbi, AJ~Banday, RB~Barreiro, JG~Bartlett, et~al.
\newblock Planck early results. xviii. the power spectrum of cosmic infrared background anisotropies.
\newblock {\em Astronomy \& Astrophysics}, 536:A18, 2011.

\bibitem{galloni2023unraveling}
Giacomo Galloni, Mario Ballardini, Nicola Bartolo, Alessandro Gruppuso, Luca Pagano, and Angelo Ricciardone.
\newblock Unraveling the cmb lack-of-correlation anomaly with the cosmological gravitational wave background.
\newblock {\em Journal of Cosmology and Astroparticle Physics}, 2023(10):013, 2023.

\bibitem{scott2016information}
Douglas Scott, Dagoberto Contreras, Ali Narimani, and Yin-Zhe Ma.
\newblock The information content of cosmic microwave background anisotropies.
\newblock {\em Journal of Cosmology and Astroparticle Physics}, 2016(06):046, 2016.

\bibitem{Planck:2019evm}
Y.~Akrami et~al.
\newblock {Planck 2018 results. VII. Isotropy and Statistics of the CMB}.
\newblock {\em Astron. Astrophys.}, 641:A7, 2020.

\bibitem{giare2023cmb}
William Giar{\`e}.
\newblock Cmb anomalies and the hubble tension.
\newblock {\em arXiv preprint arXiv:2305.16919}, 2023.

\bibitem{hu2023hubble}
Jian-Ping Hu and Fa-Yin Wang.
\newblock Hubble tension: the evidence of new physics.
\newblock {\em Universe}, 9(2):94, 2023.

\bibitem{abdalla2022cosmology}
Elcio Abdalla, Guillermo~Franco Abell{\'a}n, Amin Aboubrahim, Adriano Agnello, {\"O}zg{\"u}r Akarsu, Yashar Akrami, George Alestas, Daniel Aloni, Luca Amendola, Luis~A Anchordoqui, et~al.
\newblock Cosmology intertwined: A review of the particle physics, astrophysics, and cosmology associated with the cosmological tensions and anomalies.
\newblock {\em Journal of High Energy Astrophysics}, 34:49--211, 2022.

\bibitem{krishnan2023dipole}
Chethan Krishnan, Ranjini Mondol, and MM~Sheikh-Jabbari.
\newblock Dipole cosmology: the copernican paradigm beyond flrw.
\newblock {\em Journal of Cosmology and Astroparticle Physics}, 2023(07):020, 2023.

\bibitem{verde2013planck}
Licia Verde, Pavlos Protopapas, and Raul Jimenez.
\newblock Planck and the local universe: Quantifying the tension.
\newblock {\em Physics of the Dark Universe}, 2(3):166--175, 2013.

\bibitem{fields2011primordial}
Brian~D Fields.
\newblock The primordial lithium problem.
\newblock {\em Annual Review of Nuclear and Particle Science}, 61(1):47--68, 2011.

\bibitem{perivolaropoulos2021hubble}
Leandros Perivolaropoulos and Foteini Skara.
\newblock Hubble tension or a transition of the cepheid snia calibrator parameters?
\newblock {\em Physical Review D}, 104(12):123511, 2021.

\bibitem{riess2022comprehensive}
Adam~G Riess, Wenlong Yuan, Lucas~M Macri, Dan Scolnic, Dillon Brout, Stefano Casertano, David~O Jones, Yukei Murakami, Gagandeep~S Anand, Louise Breuval, et~al.
\newblock A comprehensive measurement of the local value of the hubble constant with 1 km s- 1 mpc- 1 uncertainty from the hubble space telescope and the sh0es team.
\newblock {\em The Astrophysical journal letters}, 934(1):L7, 2022.

\bibitem{lemos2023cosmic}
Pablo Lemos and Paul Shah.
\newblock The cosmic microwave background and $ h\_0$.
\newblock {\em arXiv preprint arXiv:2307.13083}, 2023.

\bibitem{dark2016dark}
Dark Energy~Survey Collaboration:, T~Abbott, FB~Abdalla, J~Aleksi{\'c}, S~Allam, A~Amara, D~Bacon, E~Balbinot, M~Banerji, K~Bechtol, et~al.
\newblock The dark energy survey: more than dark energy--an overview.
\newblock {\em Monthly Notices of the Royal Astronomical Society}, 460(2):1270--1299, 2016.

\bibitem{dark2005dark}
Dark Energy Survey Collaboration: T~Abbott et~al.
\newblock The dark energy survey.
\newblock {\em International Journal of Modern Physics A}, 20(14):3121--3123, 2005.

\bibitem{hwang2001f}
Jai-chan Hwang and Hyerim Noh.
\newblock f (r) gravity theory and cmbr constraints.
\newblock {\em Physics Letters B}, 506(1-2):13--19, 2001.

\bibitem{bessa2022observational}
Pedro Bessa, Marcela Campista, and Armando Bernui.
\newblock Observational constraints on starobinsky f (r) cosmology from cosmic expansion and structure growth data.
\newblock {\em The European Physical Journal C}, 82(6):506, 2022.

\bibitem{aviles2019screenings}
Alejandro Aviles, Jorge~L Cervantes-Cota, and David~F Mota.
\newblock Screenings in modified gravity: a perturbative approach.
\newblock {\em Astronomy \& Astrophysics}, 622:A62, 2019.

\bibitem{starobinsky1980new}
Alexei~A. Starobinsky.
\newblock A new type of isotropic cosmological models without singularity.
\newblock {\em Physics Letters B}, 91(1):99--102, 1980.

\bibitem{basilakos2013observational}
Spyros Basilakos, Savvas Nesseris, and Leandros Perivolaropoulos.
\newblock Observational constraints on viable $f(r)$ parametrizations with geometrical and dynamical probes.
\newblock {\em Physical Review D—Particles, Fields, Gravitation, and Cosmology}, 87(12):123529, 2013.

\bibitem{jana2019constraints}
Soumya Jana and Subhendra Mohanty.
\newblock Constraints on f (r) theories of gravity from gw170817.
\newblock {\em Physical Review D}, 99(4):044056, 2019.

\bibitem{nunes2017new}
Rafael~C Nunes, Supriya Pan, Emmanuel~N Saridakis, and Everton~MC Abreu.
\newblock New observational constraints on $f(r)$ gravity from cosmic chronometers.
\newblock {\em Journal of Cosmology and Astroparticle Physics}, 2017(01):005, 2017.

\bibitem{hu2007models}
Wayne Hu and Ignacy Sawicki.
\newblock Models of f (r) cosmic acceleration that evade solar system tests.
\newblock {\em Physical Review D—Particles, Fields, Gravitation, and Cosmology}, 76(6):064004, 2007.

\bibitem{casas2023euclid}
Santiago Casas, VF~Cardone, Domenico Sapone, N~Frusciante, Francesco Pace, Gabriele Parimbelli, M~Archidiacono, K~Koyama, Isaac Tutusaus, Stefano Camera, et~al.
\newblock Euclid: Constraints on f (r) cosmologies from the spectroscopic and photometric primary probes.
\newblock {\em arXiv preprint arXiv:2306.11053}, 2023.

\bibitem{ocampo2024enhancing}
Indira Ocampo, George Alestas, Savvas Nesseris, and Domenico Sapone.
\newblock Enhancing cosmological model selection with interpretable machine learning.
\newblock {\em arXiv preprint arXiv:2406.08351}, 2024.

\bibitem{features_review}
Jens {Chluba}, Jan {Hamann}, and Subodh~P. {Patil}.
\newblock {Features and new physical scales in primordial observables: Theory and observation}.
\newblock {\em International Journal of Modern Physics D}, 24(10):1530023, June 2015.

\bibitem{alam2017clustering}
Shadab Alam, Metin Ata, Stephen Bailey, Florian Beutler, Dmitry Bizyaev, Jonathan~A Blazek, Adam~S Bolton, Joel~R Brownstein, Angela Burden, Chia-Hsun Chuang, et~al.
\newblock The clustering of galaxies in the completed sdss-iii baryon oscillation spectroscopic survey: cosmological analysis of the dr12 galaxy sample.
\newblock {\em Monthly Notices of the Royal Astronomical Society}, 470(3):2617--2652, 2017.

\bibitem{abbott2022dark}
Timothy~MC Abbott, Michel Aguena, Alex Alarcon, S~Allam, O~Alves, A~Amon, F~Andrade-Oliveira, James Annis, S~Avila, D~Bacon, et~al.
\newblock Dark energy survey year 3 results: Cosmological constraints from galaxy clustering and weak lensing.
\newblock {\em Physical Review D}, 105(2):023520, 2022.

\bibitem{de2013kilo}
Jelte~TA de~Jong, Gijs~A Verdoes~Kleijn, Konrad~H Kuijken, Edwin~A Valentijn, KiDS, and Astro-WISE Consortiums.
\newblock The kilo-degree survey.
\newblock {\em Experimental Astronomy}, 35:25--44, 2013.

\bibitem{aiola2020atacama}
Simone Aiola, Erminia Calabrese, Lo{\"\i}c Maurin, Sigurd Naess, Benjamin~L Schmitt, Maximilian~H Abitbol, Graeme~E Addison, Peter~AR Ade, David Alonso, Mandana Amiri, et~al.
\newblock The atacama cosmology telescope: Dr4 maps and cosmological parameters.
\newblock {\em Journal of Cosmology and Astroparticle Physics}, 2020(12):047, 2020.

\bibitem{blanchard2020euclid}
Alain Blanchard, S~Camera, Carmelita Carbone, VF~Cardone, S~Casas, S{\'e}bastien Clesse, S~Ili{\'c}, M~Kilbinger, T~Kitching, Martin Kunz, et~al.
\newblock Euclid preparation-vii. forecast validation for euclid cosmological probes.
\newblock {\em Astronomy \& Astrophysics}, 642:A191, 2020.

\bibitem{laureijs2011euclid}
Rene Laureijs, J{\'e}r{\^o}me Amiaux, S~Arduini, J-L Augueres, J~Brinchmann, R~Cole, M~Cropper, C~Dabin, L~Duvet, A~Ealet, et~al.
\newblock Euclid definition study report.
\newblock {\em arXiv preprint arXiv:1110.3193}, 2011.

\bibitem{mellier2024euclid}
Y~Mellier, JA~Barroso, A~Ach{\'u}carro, J~Adamek, R~Adam, GE~Addison, N~Aghanim, M~Aguena, V~Ajani, Y~Akrami, et~al.
\newblock Euclid. i. overview of the euclid mission.
\newblock {\em arXiv preprint arXiv:2405.13491}, 2024.

\bibitem{ivezic2019lsst}
{\v{Z}}eljko Ivezi{\'c}, Steven~M Kahn, J~Anthony Tyson, Bob Abel, Emily Acosta, Robyn Allsman, David Alonso, Yusra AlSayyad, Scott~F Anderson, John Andrew, et~al.
\newblock Lsst: from science drivers to reference design and anticipated data products.
\newblock {\em The Astrophysical Journal}, 873(2):111, 2019.

\bibitem{abareshi2022overview}
Behzad Abareshi, J~Aguilar, S~Ahlen, Shadab Alam, David~M Alexander, R~Alfarsy, L~Allen, C~Allende Prieto, O~Alves, J~Ameel, et~al.
\newblock Overview of the instrumentation for the dark energy spectroscopic instrument.
\newblock {\em The Astronomical Journal}, 164(5):207, 2022.

\bibitem{joseph2021joint}
R{\'e}my Joseph, Peter Melchior, and Fred Moolekamp.
\newblock Joint survey processing: combined resampling and convolution for galaxy modelling and deblending.
\newblock {\em arXiv preprint arXiv:2107.06984}, 2021.

\bibitem{dvorkin2022machine}
Cora Dvorkin, Siddharth Mishra-Sharma, Brian Nord, V~Ashley Villar, Camille Avestruz, Keith Bechtol, Aleksandra {\'C}iprijanovi{\'c}, Andrew~J Connolly, Lehman~H Garrison, Gautham Narayan, et~al.
\newblock Machine learning and cosmology.
\newblock {\em arXiv preprint arXiv:2203.08056}, 2022.

\bibitem{moriwaki2023machine}
Kana Moriwaki, Takahiro Nishimichi, and Naoki Yoshida.
\newblock Machine learning for observational cosmology.
\newblock {\em Reports on Progress in Physics}, 86(7):076901, 2023.

\bibitem{rose2024introducing}
Jonah~C Rose, Paul Torrey, Francisco Villaescusa-Navarro, Mariangela Lisanti, Tri Nguyen, Sandip Roy, Kassidy~E Kollmann, Mark Vogelsberger, Francis-Yan Cyr-Racine, Mikhail~V Medvedev, et~al.
\newblock Introducing the dreams project: Dark matter and astrophysics with machine learning and simulations.
\newblock {\em arXiv preprint arXiv:2405.00766}, 2024.

\bibitem{murakami2023non}
Koya Murakami, Indira Ocampo, Savvas Nesseris, Atsushi~J Nishizawa, and Sachiko Kuroyanagi.
\newblock Non-linearity-free prediction of the growth-rate $f\sigma_8 $ using convolutional neural networks.
\newblock {\em arXiv preprint arXiv:2305.12812}, 2023.

\bibitem{min2024deep}
Zhiwei Min, Xu~Xiao, Jiacheng Ding, Liang Xiao, Jie Jiang, Donglin Wu, Qiufan Lin, Yin Li, Yang Wang, Shuai Liu, et~al.
\newblock Deep learning for cosmological parameter inference from dark matter halo density field.
\newblock {\em arXiv preprint arXiv:2404.09483}, 2024.

\bibitem{garcia2024bayesian}
Jorge~Enrique Garc{\'\i}a-Farieta, H{\'e}ctor~J Hort{\'u}a, and Francisco-Shu Kitaura.
\newblock Bayesian deep learning for cosmic volumes with modified gravity.
\newblock {\em Astronomy \& Astrophysics}, 684:A100, 2024.

\bibitem{peel2019distinguishing}
Austin Peel, Florian Lalande, Jean-Luc Starck, Valeria Pettorino, Julian Merten, Carlo Giocoli, Massimo Meneghetti, and Marco Baldi.
\newblock Distinguishing standard and modified gravity cosmologies with machine learning.
\newblock {\em Physical Review D}, 100(2):023508, 2019.

\bibitem{yan2023lensing}
Ye-Peng Yan, Guo-Jian Wang, Si-Yu Li, Yang-Jie Yan, and Jun-Qing Xia.
\newblock Lensing reconstruction from the cosmic microwave background polarization with machine learning.
\newblock {\em The Astrophysical Journal}, 952(1):15, 2023.

\bibitem{kamerkar2023machine}
Ahana Kamerkar, Savvas Nesseris, and Lucas Pinol.
\newblock Machine learning cosmic inflation.
\newblock {\em Physical Review D}, 108(4):043509, 2023.

\bibitem{remazeilles2011foreground}
Mathieu Remazeilles, Jacques Delabrouille, and Jean-Fran{\c{c}}ois Cardoso.
\newblock Foreground component separation with generalized internal linear combination.
\newblock {\em Monthly Notices of the Royal Astronomical Society}, 418(1):467--476, 2011.

\bibitem{wang2022recovering}
Guo-Jian Wang, Hong-Liang Shi, Ye-Peng Yan, Jun-Qing Xia, Yan-Yun Zhao, Si-Yu Li, and Jun-Feng Li.
\newblock Recovering the cmb signal with machine learning.
\newblock {\em The Astrophysical Journal Supplement Series}, 260(1):13, 2022.

\bibitem{yan2023recovering}
Ye-Peng Yan, Guo-Jian Wang, Si-Yu Li, and Jun-Qing Xia.
\newblock Recovering cosmic microwave background polarization signals with machine learning.
\newblock {\em The Astrophysical Journal}, 947(1):29, 2023.

\bibitem{rojas2020classifying}
Felipe Rojas, Lo{\"\i}c Maurin, Rolando D{\"u}nner, and Karim Pichara.
\newblock Classifying cmb time-ordered data through deep neural networks.
\newblock {\em Monthly Notices of the Royal Astronomical Society}, 494(3):3741--3749, 2020.

\bibitem{mccarthy2024signal}
Fiona McCarthy, J~Colin Hill, William~R Coulton, and David~W Hogg.
\newblock Signal-preserving cmb component separation with machine learning.
\newblock {\em arXiv preprint arXiv:2404.03557}, 2024.

\bibitem{de2022cosmic}
Mart{\'\i}n De~Los~Rios.
\newblock Cosmic-kite: auto-encoding the cosmic microwave background.
\newblock {\em Monthly Notices of the Royal Astronomical Society}, 511(4):5525--5535, 2022.

\bibitem{salti2022deep}
Mehmet Salti and Evrim~Ersin Kangal.
\newblock Deep learning of cmb radiation temperature.
\newblock {\em Annals of Physics}, 439:168799, 2022.

\bibitem{ballardini2016probing}
Mario Ballardini, Fabio Finelli, Cosimo Fedeli, and Lauro Moscardini.
\newblock Probing primordial features with future galaxy surveys.
\newblock {\em Journal of Cosmology and Astroparticle Physics}, 2016(10):041, 2016.

\bibitem{ballardini2018probing}
Mario Ballardini, Fabio Finelli, Roy Maartens, and Lauro Moscardini.
\newblock Probing primordial features with next-generation photometric and radio surveys.
\newblock {\em Journal of Cosmology and Astroparticle Physics}, 2018(04):044, 2018.

\bibitem{baumann2018tasi}
Daniel Baumann.
\newblock Tasi lectures on primordial cosmology.
\newblock {\em arXiv preprint arXiv:1807.03098}, 2018.

\bibitem{mactavish2006cmb}
Carolyn~Judith MacTavish.
\newblock {\em CMB angular power spectra and cosmological implications from the 2003 LDB flight of the BOOMERANG telescope}.
\newblock University of Toronto, 2006.

\bibitem{camb}
Antony Lewis, Anthony Challinor, and Anthony Lasenby.
\newblock Efficient computation of cosmic microwave background anisotropies in closed friedmann-robertson-walker models.
\newblock {\em The Astrophysical Journal}, 538(2):473--476, 2000.

\bibitem{class}
Julien {Lesgourgues}.
\newblock {The Cosmic Linear Anisotropy Solving System (CLASS) I: Overview}.
\newblock {\em arXiv e-prints}, page arXiv:1104.2932, April 2011.

\bibitem{kou2024constraining}
Rapha{\"e}l Kou, Calum Murray, and James~G Bartlett.
\newblock Constraining f (r) gravity with cross-correlation of galaxies and cosmic microwave background lensing.
\newblock {\em Astronomy \& Astrophysics}, 686:A193, 2024.

\bibitem{tsujikawa2007matter}
Shinji Tsujikawa.
\newblock Matter density perturbations and effective gravitational constant in modified gravity models of dark energy.
\newblock {\em Physical Review D—Particles, Fields, Gravitation, and Cosmology}, 76(2):023514, 2007.

\bibitem{motohashi2011f}
Hayato Motohashi, Alexei~A Starobinsky, and Jun'ichi Yokoyama.
\newblock f (r) gravity and its cosmological implications.
\newblock {\em International Journal of Modern Physics D}, 20(08):1347--1355, 2011.

\bibitem{song2007large}
Yong-Seon Song, Wayne Hu, and Ignacy Sawicki.
\newblock Large scale structure of f (r) gravity.
\newblock {\em Physical Review D—Particles, Fields, Gravitation, and Cosmology}, 75(4):044004, 2007.

\bibitem{ravi2024investigating}
Kumar Ravi, Anirban Chatterjee, Biswajit Jana, and Abhijit Bandyopadhyay.
\newblock Investigating the accelerated expansion of the universe through updated constraints on viable f (r) models within the metric formalism.
\newblock {\em Monthly Notices of the Royal Astronomical Society}, 527(3):7626--7651, 2024.

\bibitem{kumar2023new}
Suresh Kumar, Rafael~C Nunes, Supriya Pan, and Priya Yadav.
\newblock New late-time constraints on f (r) gravity.
\newblock {\em Physics of the Dark Universe}, 42:101281, 2023.

\bibitem{wang2022pantheon+}
Deng Wang.
\newblock Pantheon+ constraints on dark energy and modified gravity: An evidence of dynamical dark energy.
\newblock {\em Physical Review D}, 106(6):063515, 2022.

\bibitem{lodha2023searching}
Kushal Lodha, Lucas Pinol, Savvas Nesseris, Arman Shafieloo, Wuhyun Sohn, and Matteo Fasiello.
\newblock Searching for local features in primordial power spectrum using genetic algorithms.
\newblock {\em arXiv preprint arXiv:2308.04940}, 2023.

\bibitem{ballardini2024euclid}
M~Ballardini, Y~Akrami, F~Finelli, D~Karagiannis, Baojiu Li, Y~Li, Z~Sakr, D~Sapone, A~Ach{\'u}carro, M~Baldi, et~al.
\newblock Euclid: The search for primordial features.
\newblock {\em Astronomy \& Astrophysics}, 683:A220, 2024.

\bibitem{schmidhuber2015deep}
J{\"u}rgen Schmidhuber.
\newblock Deep learning in neural networks: An overview.
\newblock {\em Neural networks}, 61:85--117, 2015.

\bibitem{sakr2022cosmological}
Ziad Sakr and Matteo Martinelli.
\newblock Cosmological constraints on sub-horizon scales modified gravity theories with mgclass ii.
\newblock {\em Journal of Cosmology and Astroparticle Physics}, 2022(05):030, 2022.

\bibitem{patro2015normalization}
SGOPAL Patro and Kishore~Kumar Sahu.
\newblock Normalization: A preprocessing stage.
\newblock {\em arXiv preprint arXiv:1503.06462}, 2015.

\bibitem{agarap2018deep}
Abien~Fred Agarap.
\newblock Deep learning using rectified linear units (relu).
\newblock {\em arXiv preprint arXiv:1803.08375}, 2018.

\bibitem{srivastava2014dropout}
Nitish Srivastava, Geoffrey Hinton, Alex Krizhevsky, Ilya Sutskever, and Ruslan Salakhutdinov.
\newblock Dropout: a simple way to prevent neural networks from overfitting.
\newblock {\em The journal of machine learning research}, 15(1):1929--1958, 2014.

\bibitem{bellini2018comparison}
E~Bellini, A~Barreira, N~Frusciante, B~Hu, S~Peirone, M~Raveri, Miguel Zumalacarregui, A~Avilez-Lopez, M~Ballardini, RA~Battye, et~al.
\newblock Comparison of einstein-boltzmann solvers for testing general relativity.
\newblock {\em Physical Review D}, 97(2):023520, 2018.

\bibitem{zhang2021survey}
Yu~Zhang, Peter Ti{\v{n}}o, Ale{\v{s}} Leonardis, and Ke~Tang.
\newblock A survey on neural network interpretability.
\newblock {\em IEEE Transactions on Emerging Topics in Computational Intelligence}, 5(5):726--742, 2021.

\bibitem{lundberg2020local}
Scott~M Lundberg, Gabriel Erion, Hugh Chen, Alex DeGrave, Jordan~M Prutkin, Bala Nair, Ronit Katz, Jonathan Himmelfarb, Nisha Bansal, and Su-In Lee.
\newblock From local explanations to global understanding with explainable ai for trees.
\newblock {\em Nature machine intelligence}, 2(1):56--67, 2020.

\bibitem{sundararajan2020many}
Mukund Sundararajan and Amir Najmi.
\newblock The many shapley values for model explanation.
\newblock In {\em International conference on machine learning}, pages 9269--9278. PMLR, 2020.

\bibitem{ghorbani2019data}
Amirata Ghorbani and James Zou.
\newblock Data shapley: Equitable valuation of data for machine learning.
\newblock In {\em International conference on machine learning}, pages 2242--2251. PMLR, 2019.

\bibitem{Valle:2020jyk}
Luis Fernando Machado~Poletti Valle, Camille Avestruz, David~J. Barnes, Arya Farahi, Erwin~T. Lau, and Daisuke Nagai.
\newblock {shaping the gas: understanding gas shapes in dark matter haloes with interpretable machine learning}.
\newblock {\em Mon. Not. Roy. Astron. Soc.}, 507(1):1468--1484, 2021.

\bibitem{SHAPphotometry}
S.~{Gilda}, S.~{Lower}, and D.~{Narayanan}.
\newblock {mirkwood: Fast and Accurate SED Modeling Using Machine Learning}.
\newblock In {\em American Astronomical Society Meeting Abstracts}, volume 237 of {\em American Astronomical Society Meeting Abstracts}, page 215.07D, January 2021.

\end{thebibliography}

\end{document}